\documentclass[a4paper,11pt]{article}
\pdfoutput=1 
\usepackage{jheppub}
\usepackage{verbatim,comment}
\usepackage{ulem}
\usepackage[T1]{fontenc}
\usepackage{bm}
\usepackage{xcolor}
\usepackage{amsmath}
\usepackage{qcircuit}
\usepackage{float}
\usepackage{braket}
\usepackage{booktabs}
\usepackage{subfigure}
\usepackage{hyperref}
\usepackage[us,12hr]{datetime}
\usepackage{braket}
\usepackage[toc,page]{appendix}

\def\be{\begin{equation}}
\def\ee{\end{equation}}
\def\bea{\begin{eqnarray}}
\def\eea{\end{eqnarray}}
\newcommand{\SF}[1]{\textsc{Strawberry Fields}{#1}}

\newcommand*{\UTK}{Department of Physics and Astronomy, University of Tennessee, Knoxville, Tennessee 37996-1200, USA}

\preprint{JLAB-THY-23-3934}

\title{Continuous variable quantum computation of the $\text{O(3)}$ model in 1+1 dimensions}
\author[a]{Raghav G. Jha,}
\author[a,b]{Felix Ringer,}
\author[c]{George Siopsis,}
\author[c]{Shane Thompson}
\affiliation[a]{Thomas Jefferson National Accelerator Facility, Newport News, VA 23606, USA}
\affiliation[b]{Department of Physics, Old Dominion University, Norfolk, VA 23529, USA}
\affiliation[c]{\UTK}

\emailAdd{raghav.govind.jha@gmail.com}
\emailAdd{fmringer@jlab.org}
\emailAdd{siopsis@tennessee.edu}
\emailAdd{sthomp78@tennessee.edu}

\abstract{We formulate the $\text{O(3)}$ non-linear sigma model in 1+1 dimensions as a limit of a three-component scalar field theory restricted to the unit sphere in the large squeezing limit. This allows us to describe the model in terms of the continuous variable (CV) approach to quantum computing. We construct the ground state and excited states using the coupled-cluster Ansatz and find excellent agreement with the exact diagonalization results for a small number of lattice sites. We then present the simulation protocol for the time evolution of the model using CV gates and obtain numerical results using a photonic quantum simulator. We expect that the methods developed in this work will be useful for exploring interesting dynamics for a wide class of sigma models and gauge theories, as well as for simulating scattering events on quantum hardware in the coming decades.}

\begin{document} 
\maketitle
\flushbottom

\section{Introduction}

All fundamental interactions in nature except gravity have been successfully described within the framework of quantum field theory. A full understanding of the real-time dynamics of these interacting field theories is still an open problem. In lower dimensions, some progress has been made using classical computations based on MPS/PEPS tensor network methods \cite{RevModPhys.93.045003}.  
However, for theories close to the critical point these methods are limited because they can only efficiently represent ground states of a special class of local Hamiltonians with a gapped spectrum due to their 
peculiar entanglement scaling. Another limitation is that even for systems with a gapped spectrum, the growth of entanglement during real-time evolution can render these methods ineffective. The limitations of these classical methods are expected to be overcome by quantum computers, which has led to an increased effort aimed at understanding various lattice models and lattice field theories using quantum computing algorithms.

From the perspective of fundamental nuclear and particle physics~\cite{Bauer:2022hpo}, the long-term goal is the study of real-time dynamics of quantum chromodynamics (QCD) in four dimensions, which is relevant for the description of inelastic scattering processes at current and future collider experiments~\cite{Briceno:2020rar}. Due to current hardware limitations, lower-dimensional models have typically been considered that represent important stepping stones toward simulations of QCD. In this regard, the $\text{O(3)}$ nonlinear sigma model in 1+1 dimensions is a particularly interesting test ground. This model exhibits a global non-Abelian symmetry group and shares several interesting properties with QCD, the most important of which is asymptotic freedom~\cite{Polyakov:1975rr}. The $\text{O(3)}$ model has a dynamical mass gap and also admits instanton solutions similar to four-dimensional QCD~\cite{Polyakov:1975yp}.
In addition, it is relevant for low-energy dynamics of pions in nuclear physics as discussed in more detail below. This makes it a preferred toy model~\cite{Berg:1981er} to explore fundamental questions related to nuclear and particle physics. Even though this model has been extensively studied with classical numerical methods, it has recently attracted increased interest because new classical and quantum computing methods have been developed. This model has been studied using tensor network techniques based on matrix product states (MPS) with and without a topological term (for $\theta =  \pi,0$)~\cite{Bruckmann:2018usp, Tang:2021uge} and using higher-order tensor renormalization group methods~\cite{Unmuth-Yockey:2014afa}. Different regularizations for lattice simulations of the $\text{O(3)}$ model have been explored such as the fuzzy sphere qubitization, D-theory, the angular momentum basis, the Heisenberg comb, and the Schwinger boson formulation~\cite{Brower:2003vy,Chandrasekharan:2001ya,Singh:2019jog,Alexandru:2019ozf,Buser:2020uzs,Caspar:2022llo,Araz:2022tbd,Alexandru:2022son,Jha:2023ump}. In Ref.~\cite{Ciavarella:2022qdx}, simulations and the preparation of ground states of the $\text{O(3)}$ model were discussed in the context of cold atom quantum simulators using near-term quantum platforms. In order to regularize the theory and carry out numerical computations, recent efforts have mostly focused on the qubit approach to quantum computing~\cite{Singh:2019uwd, Alexandru:2021xkf,Araz:2022tbd,Ciavarella:2022qdx}. Within the qubitization program, it has been argued that to reproduce the critical point of the continuum field theory in this model, only two qubits per site are required. 

In this work, we present an alternative method, which was been argued to be more natural than qubit-based quantum computing~\cite{Jordan:2011ci} in simulating, for example, bosonic models or lattice field theories such as $\phi^4$ scalar field theory~\cite{Marshall:2015mna,Yeter2022}. This approach is known as continuous variable (CV) quantum computing~\cite{PhysRevLett.82.1784}, which we employ in this work for simulations of the $\text{O(3)}$ model. Instead of qubits, the fundamental unit to carry out computations is qumodes which can be represented by quantum mechanical harmonic oscillators. The fundamental idea of the CV approach is to not consider an array of two-level systems such as qubits or $d$-state systems such as qudits but to harness the power of the infinite-dimensional representation in terms of bosonic fields obeying the infinite-dimensional commutator relations. Qumodes may even be used as a resource for qubit many-body calculations, see Ref.\ \cite{PhysRevLett.127.020502} for an application to finite-temperature systems. While continuous variable quantum computations can be realized with different hardware platforms, we will primarily focus on photonic systems~\cite{PhysRevLett.97.110501}. Recent technological advances make photonic platforms a promising candidate for scalable quantum computations. For example, Ref.~\cite{Xanadu22} reported a quantum computational advantage with a programmable photonic processor in the context of Gaussian Boson Sampling with 216 squeezed qumodes. Additionally, the realization of a 20-qumode universal quantum photonic processor where unitary operations can be performed with high fidelity was described in Ref.~\cite{Taballione:2022xjq}. In Ref.~\cite{Eaton:2022vjq}, photon number resolution measurements of up to 100 photons have been developed by sampling from a coherent state with increased robustness against environmental noise. 

The outline of the paper is as follows. In Sec.~\ref{sec:descr}, we start by discussing the relevance of the $\text{O(3)}$ model for fundamental nuclear and particle physics. We then review the standard rotor Hamiltonian which describes the lattice version of the sigma model in one spatial dimension, and consider the coupled-cluster Ansatz for the ground and first excited states of the $\text{O(3)}$ model. We then present the CV formulation in Sec.~\ref{sec:qumode}. In Sec.~\ref{sec:quantum_sim},  we present the protocol to measure energy expectation values and perform time evolution of the model using the \textsc{Strawberry Fields}~\cite{Killoran_2019} quantum simulator. We end the paper with a summary and conclusions in Sec.~\ref{sec:conclu}. We provide additional details about the CV gates in Appendix \ref{sec:CV_gates}. 

\section{The \text{O(3)} nonlinear sigma model - a toy model for nuclear physics~\label{sec:descr}}

Nonlinear sigma models have been extensively studied because they share several features with gauge theories but without added complications related to maintaining gauge invariance. An important application of these models is in the low-energy dynamics of pions described by an effective chiral Lagrangian density given schematically by $\mathcal{L} = \frac{1}{4} \text{Tr} [ \partial_\mu U \partial^\mu U^\dagger]$, where $U$ is an isospin $\text{SU(2)}$ matrix. Due to this reason, this is also sometimes referred to as a ``principal chiral field'' model
\cite{Polyakov:1987ez}. It has a global $\text{SU(2)}_L \times \text{SU(2)}_R$ symmetry which coincides with the $O(4)$ symmetry of the sigma model. This is clearly seen if we parametrize the isospin matrix as $U = n_0 \mathbb{I} + i\bm{n}\cdot \bm{\sigma}$, where $\bm{\sigma} = (\sigma_1,\sigma_2,\sigma_3)$ are Pauli matrices, and $n_an^a \equiv n_0^2 + \bm{n}^2 = 1$. We can define the angular momenta as $J_{ab} = -i \big(n_a \frac{\partial}{\partial n_b} - n_b \frac{\partial}{\partial n_a}\big)$ with $a,b=0,1,2,3$. The Hamiltonian discretized on a spatial lattice can be written as \be {H} = \frac{1}{2g^2} \sum_{a,b} J_{ab}^2 - g^2 \sum_{\langle x,x^{\prime} \rangle} n_a(x) n^a (x') \ee where $g$ is a coupling constant. The potential is bi-linear in the vectors $n_a$ that act as coordinates of the system.

Moreover, a gauge theory with a local $\text{SU(2)}$ symmetry can also be formulated in terms of the parametrized isospin matrix on the lattice. In this case, the matrices $U$ reside on the links along which one also defines angular momenta $J_{ab}$. The Hamiltonian can be written as
\be {H} = \frac{1}{2g^2} \sum_{\text{links}} J_{ab}^2 - \frac{g^2}{2} \sum_{\text{plaquettes}} \text{Tr} [U(1) U(2) U(3) U(4)] \ee 
where we introduced the Wilson loop over a plaquette in the second term. It can be expressed in terms of the four-dimensional vectors $n_a (i)$ with $i=1,2,3,4$. We obtain a quadri-linear expression for the plaquette term
\bea \frac{1}{2}\text{Tr} [U(1) U(2) U(3) U(4)] &=& {n}^a(1)  {n}_a(2) \ {n}^b(3) {n}_b(4) - {n}^a (1){n}_a (3) \ {n}^b(2) {n}_b(4) \nonumber\\ &&+ {n}^a (1) {n}_a (4) \ {n}^b(2) {n}_b(3) +\epsilon^{abcd} n_a (1) n_b (2) n_c (3) n_d (4) \,.\;\;
\eea 
The states obey the constraint $\epsilon^{abcd} J_{ab} J_{cd} \ket{\Psi} = 0$.  Additionally, the system obeys Gauss's Law which further constrains the Hilbert space to the gauge singlet sector \cite{PhysRevD.11.395}.
However, in this work, we will focus on quantum computations using continuous variables for a simpler system, the nonlinear $\text{O(3)}$ sigma model, which shares important features with theories relevant to understanding strong interactions but is easier to tackle. We leave extensions of the approach taken in this paper for the continuous variable formulation of gauge theories for future works. The nearest-neighbor $\text{O(3)}$ sigma model Hamiltonian is given by~\cite{Hamer:1978ew}
\begin{equation}
\label{eq:Ham0}
    {H} = \frac{1}{2g^2} \sum_{i} \bm{L}^{2}_{i} - g^2 \sum_{\langle i,j \rangle} \bm{n}_{i} \cdot \bm{n}_{j} \,.
\end{equation}
Here $g^2$ is the coupling constant, $i$ and $j$ index nearest neighbor sites on a uni-directional lattice. In addition, $\bm{n}_i$ is a unit 3-vector at site $i$, which takes values on $\mathbb{S}_{2}$ and $\bm{L}_i$ is the angular momentum operator at each site, $L^a = \frac{1}{2} \epsilon^{abc} J_{bc}$. We use periodic boundary conditions. As is customary, we write the interaction term in terms of spherical coordinates noting that the vectors $\bm{n}_i$ have unit modulus:
\begin{equation}
\label{eq:n1_n2}
    \bm{n}_{i} \cdot \bm{n}_{j} = \sin \theta_i \sin\theta_j \cos(\phi_i - \phi_j) + \cos\theta_i \cos\theta_j~.
\end{equation}
In fact, one can express the dot product of the vectors in terms of spherical harmonics $Y_{l,m}(\theta, \phi)$ as: 
\begin{equation}
\label{eq:ninj_3Ys}
 \bm{n}_{i} \cdot \bm{n}_{j}  = \frac{4\pi}{3} \Big(Y_{1,0}(\theta_i, \phi_i)Y_{1,0}(\theta_j, \phi_j) - Y_{1,1}(\theta_i, \phi_i)Y_{1,-1}(\theta_j, \phi_j) -Y_{1,-1}(\theta_i, \phi_i)Y_{1,1}(\theta_j, \phi_j)
 \Big).
\end{equation} 
Each term in \eqref{eq:ninj_3Ys} has a total magnetic quantum number equal to zero. 
We can also write the interaction term in terms of $n^{\pm} = (n^{x} \pm i n^{y})/\sqrt{2}$
as \cite{Hamer:1978ew}
\begin{equation}
\label{eq:n+n-}
   \bm{n}_{i} \cdot \bm{n}_{j} =
   n^{+}_{i} n^{-}_{j} + n^{-}_{i} n^{+}_{j} + n^{z}_{i}n^{z}_{j} \,.
\end{equation}
Therefore, the lattice Hamiltonian of the 1+1-dimensional $\text{O(3)}$ model can be written as:
\begin{equation}
\label{eq:Ham_general} 
    {H} = \frac{1}{2g^2} \sum_{x=0}^{L-1} \bm{L}^{2} ({x}) - 
    g^2 \sum_{x=0}^{L-1} \Big(n^{+} ({x}) n^{-} ({x+1}) + n^{-} ({x}) n^{+} ({x+1}) + n^{z} ({x}) n^{z} ({x+1}) \Big),  
\end{equation}
and the continuum limit is obtained as we take $g^{2} \to \infty$. The eigenvalues of the kinetic term are proportional to $l(l+1)$ where $l = 0,1, \dots \infty$
denote the energy levels based on the irreducible representations of the $\text{O(3)}$ symmetry. However, for practical calculations, we impose a cutoff, which we refer to as $l_{\text{max.}}$. If we identify
$n^{\pm} = \mp X_{\pm 1}$ and $n^z = X_{0}$,
then the matrix elements of $\bm{n}$ can be computed using the well-known expressions involving two Wigner-3$j$ symbols as \cite{Wu1977}:
\begin{equation}
\langle l_1,m_1 \vert X_{M} \vert l_2, m_2 \rangle = 
(-1)^{m_1} \sqrt{(2l_1 + 1)(2l_2 + 1)}
    \begin{pmatrix}
  l_1 & 1 & l_2 \\
  0 & 0 & 0 
 \end{pmatrix}
  \begin{pmatrix}
  l_1 & 1 & l_2 \\
  -m_1 & M & m_2 
 \end{pmatrix}\, .
\end{equation}
This result is obtained from the relation
\begin{align}
     \langle l_1,m_1 \vert X_{M} \vert l_2, m_2  \rangle & = \sqrt{\frac{4\pi}{3}} \int d \Omega~Y^{*}_{l,m}Y_{1,M}Y_{l^\prime,m^\prime} \nonumber\\
    &= (-1)^m \sqrt{\frac{4\pi}{3}} \int d \Omega~Y_{l,-m} Y_{1,M}Y_{l^\prime,m^\prime}~, 
\end{align}
and the Gaunt coefficients 
\begin{equation} 
\int d \Omega~Y_{l_1,m_1}Y_{l_2,m_2}Y_{l_3,m_3}
= 
\sqrt{\frac{(2l_1 + 1)(2l_2 + 1)(2l_3 + 1)}{4\pi}}
\begin{pmatrix}
  l_1 & l_2 & l_3 \\
  0 & 0 & 0 
 \end{pmatrix}
  \begin{pmatrix}
  l_1 & l_2 & l_2 \\
  m_1 & m_2 & m_2 
 \end{pmatrix}~.
\end{equation}
In order to construct a reliable Ansatz for the $\text{O(3)}$ model, we use the coupled-cluster (CC) method which involves a state of the form \cite{Crawford2000}
\begin{equation}
\ket{\psi} \propto e^{\alpha \hat{T}} \ket{\psi_0}
\end{equation}
where $\hat{T}$ is the cluster operator built from single interaction terms in the Hamiltonian and $\alpha$ is a tunable parameter. We can include higher-interaction terms, known as double-excitation terms in the CC ansatz, but for our purposes, an operator bilinear in $\bm{n}$ suffices. The use of a CC Ansatz for sigma models is not new and this has already been explored several decades ago for a class of $O(N)$ non-linear sigma models~\cite{Kutzelnigg1991, Ligterink:1997he,Ligterink1998}. In addition, unitary CC Ans\"atze have been used for quantum computations in different contexts \cite{Ryabinkin2018,Dumitrescu:2018njn}. The unitarity or non-unitarity of the CC Ansatz depends on whether the $\hat{T}$ is anti-Hermitian or Hermitian. 
To build the CC state in our case, we start with the tensor product of the states $\ket{l=0,m=0}$ defined at each site
\be \label{eq:CCOmega0}
\ket{\Omega_0} \equiv \bigotimes_{x=0}^{L-1} \ket{l(x)=0,m(x)=0} \,, 
\ee
which is the weak coupling vacuum state corresponding to the vanishing cluster operator $\hat{T} = 0$. We choose to express $\hat{T}$ in terms of the potential term of the Hamiltonian and we obtain the following coupled-cluster Ansatz:
\be \label{eq:CC_0}\ket{\text{CC}} \propto \prod_x e^{\frac{\alpha g^2}{L} \bm{n}(x) \cdot\bm{n} (x+1)} \ket{\Omega_0}\ , \ee
where $\alpha$ is a variational parameter. As shown in Refs. \cite{Duncan1985,PhysRevD.32.3277}, this Ansatz appears to perform quite well for various coupling strengths. However, we expect that additional terms, such as terms bi-linear in $n_i$ that are not nearest-neighbor terms, as well as quadri-linear terms, which would correspond to the ``doubles excitations'' of coupled-cluster theory, would improve energy estimates for larger lattices as well as coupling $g^2$. The effectiveness of such additional terms is the subject of ongoing and future work.

For $L=2$ lattice sites, the energy of the CC state can be computed analytically. It is given  by
\be \label{eq:Leq2_analytic} 
\frac{E_0 (\alpha)}{L}  = \frac{E_0 (\alpha)}{2}  = -\frac{1}{4g^2} + \frac{1}{2\alpha} + 
 \frac{\alpha-2 g^2}{2} \coth(2 g^2 \alpha), \ee
For small $g^2$, this expression is minimized for $\alpha =0$, and we obtain the estimate of the ground state energy $E_0/2 = 0$. This corresponds to the energy of the state with zero angular momentum, as expected. Next, we observe that the optimal value of $\alpha$ goes to 1 for large values of $g^2$. This provides the following estimate of the ground state energy \be\frac{E_0}{2} \approx -g^2 + 1\ .\ee This is an expected result because at large $g^2$, the potential energy dominates and it is minimized when all unit vectors $\bm{n} (x)$ align (since $\bm{n} (x) \cdot \bm{n} (x+1) \le 1$).

 \begin{figure}
     \centering
     \subfigure[$E_0$]{\includegraphics[width=0.45\linewidth]{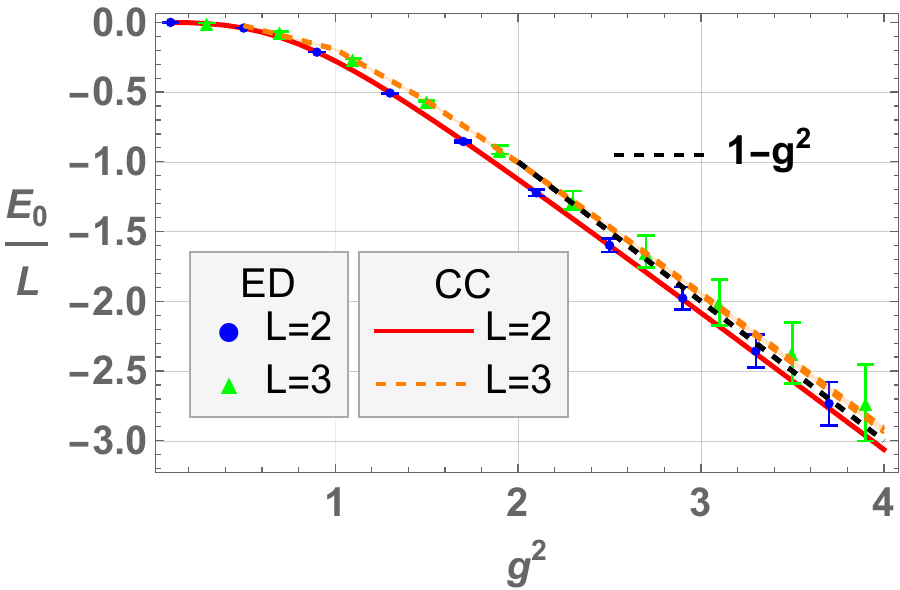}}
    \hspace{1em}
    \subfigure[$\Delta E$]{\includegraphics[width=0.45\linewidth]{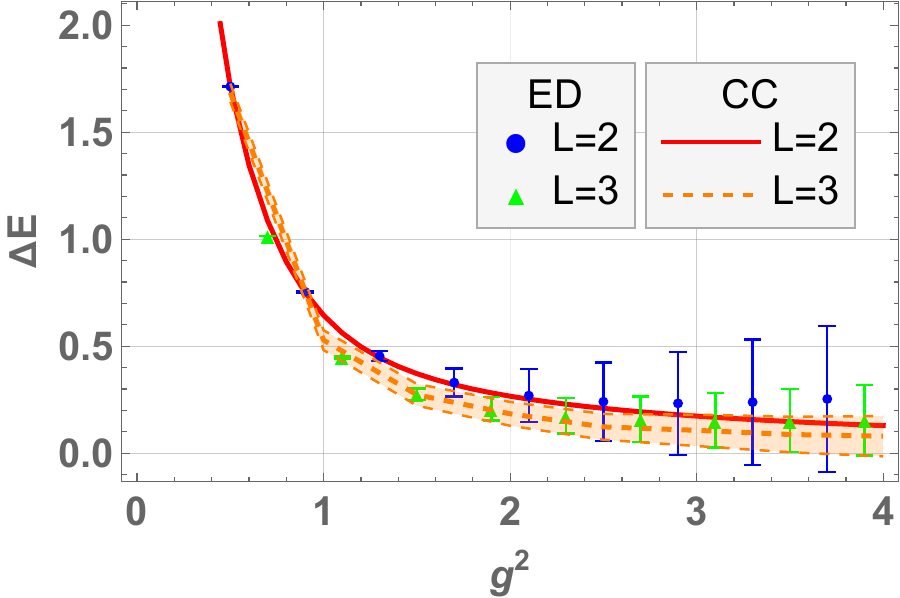}}
     \caption{The ground state energy $E_0$ and mass gap $\Delta E = E_1 - E_0$ using ED and CC for lattice sizes $L=2,3$. Angular momentum cutoff set at $l_\text{max.}=3$ for ED, with error bars given by difference from $l_\text{max.}=2$ result. 500,000 sample points for MC integration for $L=3$. $L=2$ CC is analytic. }
     \label{fig:E0}
 \end{figure}

A similar approach based on a CC Ansatz which is a modified version of \eqref{eq:CC_0} can be used to
estimate excited state energies. For the first excited state, we apply the CC Ansatz:
\be \label{eq:fe_cc0} \ket{\text{CC}_{1}} \propto e^{\frac{\alpha g^2}{L} \sum_x \bm{n}(x) \cdot\bm{n} (x+1)} \sum_x n_3(x) \ket{\Omega_0} \ee
Since $n_3 = \cos\theta$, the inclusion of $n_3(x)$ replaces $Y_{00} (\theta,\phi)$ with $Y_{10} (\theta,\phi)$ at lattice site $x$, thus introducing an angular momentum excitation of $l=1$ there.
The resulting state is orthogonal to the CC Ansatz for the ground state $\braket{\text{CC} \vert \text{CC}_1} = 0$. For illustration purposes, we consider again only two lattice sites $L=2$. The Ansatz in Eq.~\eqref{eq:fe_cc0} yields the following estimate for the energy of the first excited state
\be\label{eq:E1alpha}
\frac{E_1(\alpha)}{L}= -g^2 + \frac{-\alpha + 4 g^2 (1 + 2 g^2 \alpha - \alpha^2) + 
 e^{4 g^2 \alpha} (-4 g^2 + \alpha + 
    8 g^4 (\alpha + \alpha^3))}{4 g^2 \alpha (1+ 
   e^{4 g^2 \alpha}  (-1 + 4 g^2 \alpha))} \,.
\ee
For small $g^2$, this is minimized again for $\alpha =0$ and we obtain the following estimate of the first excited state $\frac{E_1}{2} = \frac{1}{2g^2}$. This result corresponds to a single $l=1$ excitation, as expected. In the large $g^2$ limit, the expression in Eq.~\eqref{eq:E1alpha} is minimized for $\alpha=1$, similar to the ground state, and the corresponding energy is $\frac{E_1}{2} = -g^2+1$. This demonstrates that the mass gap closes asymptotically as $g^2 \to\infty$ which is consistent with the known results.
 
The ground-state energy obtained using the CC state and the Hamiltonian \eqref{eq:Ham0} is shown in Figure \ref{fig:E0}(a). We plot $E_0(\alpha)/L$ as a function of $g^2$ for $L=2$ and $L=3$ compared to the results from exact diagonalization (ED). Using a cutoff $l_\text{max.} = 3$, for small $g^2$, we see excellent agreement between the CC and ED results up to $g^2 \approx 4$. We also find that in both methods, the uncertainty in our results increases with $g^2$. In exact diagonalization, this is because the angular momentum truncation error increases with $g^2$. It follows that the difference between the results corresponding to $l_\text{max.}-1$ and $l_\text{max.}$, which we take as our error estimate, increases as well. For the coupled-cluster method, where the $L=3$ results were computed using quasi-Monte Carlo integration \cite{Niederreiter1978}, we expect the wave function to deviate further from a uniform configuration making sampling from a uniform distribution less effective and increasing uncertainty.
The mass gaps $\Delta E=E_1-E_0$ obtained from the CC and ED methods are shown in Figure \ref{fig:E0}(b). Once again, we observe good agreement as well as a more noticeable increase in uncertainty as the value of $g^2$ is increased.

It is possible to write the $\text{O(3)}$ Hamiltonian given by \eqref{eq:Ham_general} in terms of bosonic creation and annihilation operators at each site using the Schwinger boson formalism \cite{Jha:2023ump}. This requires two qumodes for each lattice site to simulate using a quantum computation based on continuous variables. However, this approach appears challenging to implement with near- to intermediate-term resources.

In this work, we take an alternative point of view and express the $\text{O(3)}$ non-linear sigma model as a limit of a three-component scalar field theory in 1+1 dimensions requiring three qumodes at each site. We show that in the appropriate limit, this is equivalent to the rotor Hamiltonian \eqref{eq:Ham0} which is known to reproduce the sigma model in the continuum. Both of these continuous variable approaches to the $\text{O(3)}$ model belong to the same universality class. One advantage of using the approach presented here is that scalar field theories can be simulated with established methods using continuous variable quantum computation \cite{Marshall:2015mna,Thompson:2023kxz}. 

\section{Scalar field theory formulation on the 2-sphere \label{sec:qumode}}

We consider a linear $\text{O(3)}$ model consisting of real scalar fields $\phi^a(x)$ ($a=1,2,3$) in a single spatial dimension denoted by $x$. We discretize space using $L$ lattice points $x=0,1,\dots, L-1$, impose periodic boundary conditions, and choose units such that the lattice spacing, and the fundamental constants $c, \hbar$ are all set to unity. Let $\pi^a$ be the conjugate momentum to $\phi^a$ obeying the canonical commutation relations
\be [\phi^a (x) , \pi^b (x') ] = i\delta^{ab}\delta_{xx'}\, . \ee
We introduce the Hamiltonian
\be\label{eq:og_H} H= \frac{1}{2g^2} \sum_x \bm{L}^2(x) - \sum_x\left[\frac{1}{2}\left(\bm{\phi} (x) -\bm{\phi} (x+1) \right)^2-g^2\right]\,. \ee
Here the angular momentum operator $\bm{L}$ denotes the cross product of the vector field and its conjugate momentum,
\be \bm{L} (x) = \bm{\phi} (x) \times \bm{\pi} (x), \ee
where we used the simplified triplet notation for the fields, $\bm{\phi} = (\phi_1,\phi_2,\phi_3)$, and similarly for $\bm{\pi}$. 

To make contact with the nonlinear $\text{O(3)}$ model discussed in Section \ref{sec:descr}, it is useful to define a local basis at each lattice site consisting of the states defined on a three-dimensional space
\be\label{eq:basis} \ket{l,m ;\Lambda}= \frac{1}{\sqrt{\mathcal{N}}} \int dr r^2\, d^2 \bm{n}\  e^{-\frac{\Lambda^2}{8g^2} (r^2-g^2)^2} Y_{lm}\left(\bm{n}\right)\ket{r,\bm{n}}\,.
\ee
Here $\bm{n}$ is a unit 3-vector, $r$ is the radial direction, and $\Lambda$ is a (radial) momentum cutoff scale restricting the wave function on a sphere of radius $r=g$. We will model the cutoff in terms of a squeezing parameter.
The state described in~(\ref{eq:basis}) can be separated into radial and angular parts denoted by $Y_{lm}$. The radial integrand is not unique and any function that converges to a $\delta$-function centered at $r=g$ as the cutoff $\Lambda\to\infty$ would work in practice. 

In what follows, this choice affects energies only to $\mathcal{O}\left(\Lambda^{-1}\right)$.
The states in \eqref{eq:basis} form an orthonormal set with norm
\be\label{eq:35} \mathcal{N}=\frac{g^2 \sqrt{\pi}}{\Lambda}+ \mathcal{O} (\Lambda^{-3})\ . \ee
The matrix elements of the kinetic energy are given by exact expressions for finite values of the cutoff $\Lambda$ since the radial part of the wave functions decouples
\be \frac{1}{2g^2}\bra{l,m ;\Lambda} \bm{L}^2 \ket{l',m' ;\Lambda} = \frac{l(l+1)}{2g^2} \delta_{l,l'}\delta_{m,m'}\, . 
\ee
For the interaction term of the Hamiltonian, we consider two adjacent sites labeled by $i$ and $j$. We start with the matrix elements of a local term contributing to the interaction potential as
\be \bra{l,m; \Lambda} \bm{\phi}^2 \ket{l',m'; \Lambda} = g^2 \left( 1+ \mathcal{O}\left(\Lambda^{-2} \right) \right) \delta_{l,l'}\delta_{m,m'}\ . \ee
The matrix elements have a simple structure and, after subtracting $g^2$, see (\ref{eq:og_H}), they vanish in the limit of large values for the cutoff parameter $\Lambda \to\infty$.
The matrix elements of the term involving two sites can be written as
\bea 
&& \bra{l_i,m_i; \Lambda} \bra{l_j,m_j; \Lambda} \bm{\phi}(x_i)\cdot\bm{\phi}(x_j) \ket{l_i',m_i';\Lambda} \ket{l_j',m_j';\Lambda} \nonumber\\ 
= && g^2\left( 1+ \mathcal{O}\left(\Lambda^{-2}\right) \right) \bra{l_i,m_i} \bra{l_j,m_j}  \bm{n}_i\cdot\bm{n}_j \ket{l_i',m_i'} \ket{l_j',m_j'} \, . \eea 
It has the same structure as the potential term of the $\text{O(3)}$ model discussed in the previous section, which we therefore recover in the limit $\Lambda\to\infty$.

\begin{figure}
    \centering
    \subfigure[$L=2$]{\includegraphics[scale=0.5]{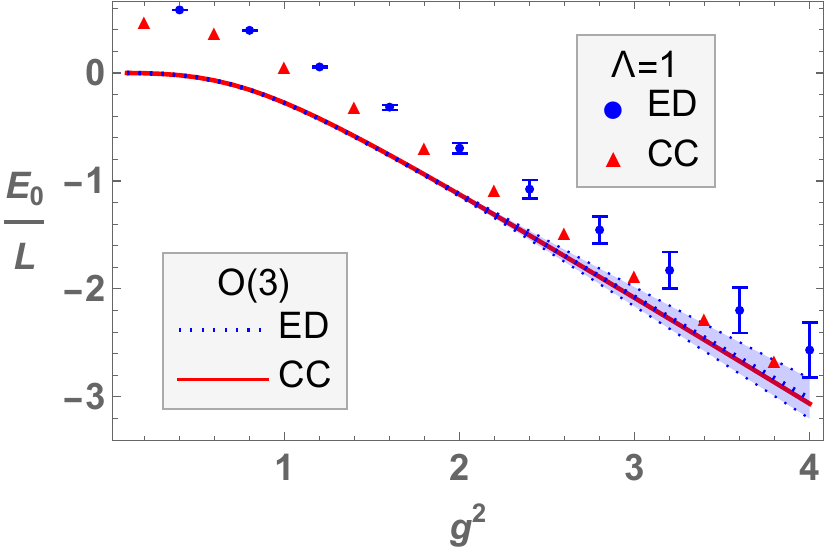}}
    \subfigure[$L=3$]{\includegraphics[scale=0.5]{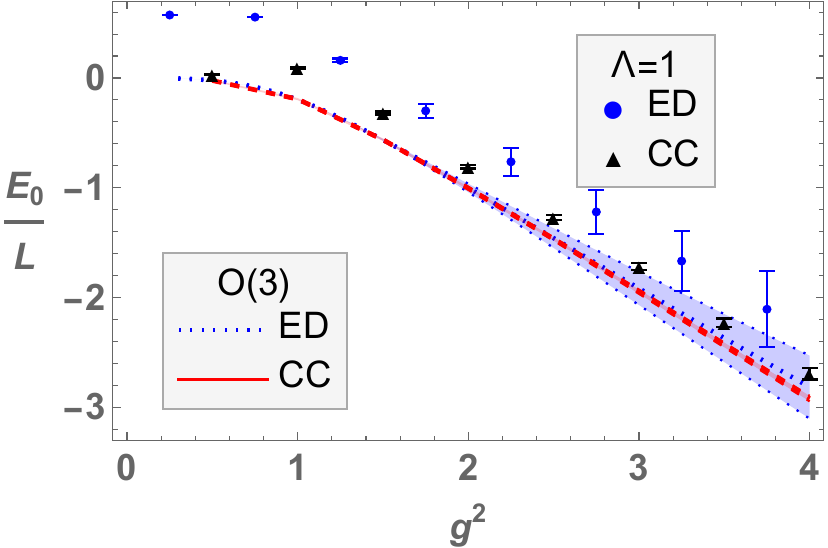}}
    \subfigure[$L=2$]{\includegraphics[scale=0.5]{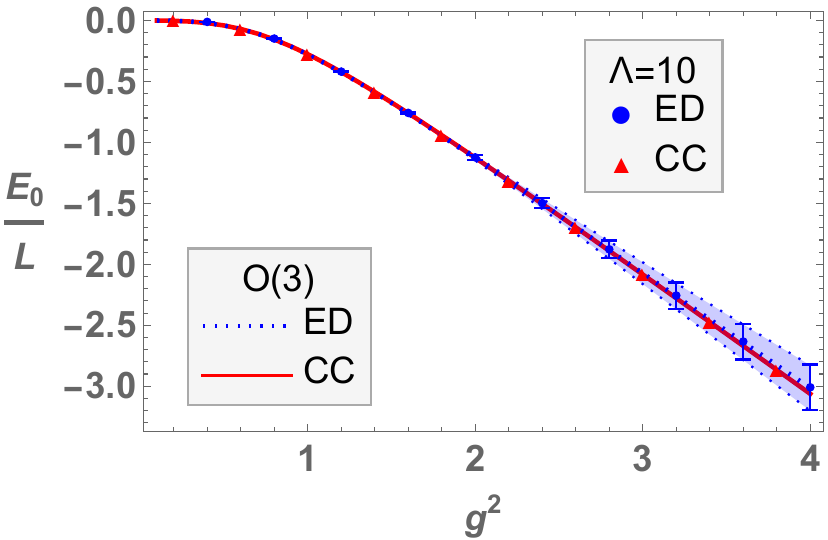}}
    \subfigure[$L=3$]{\includegraphics[scale=0.5]{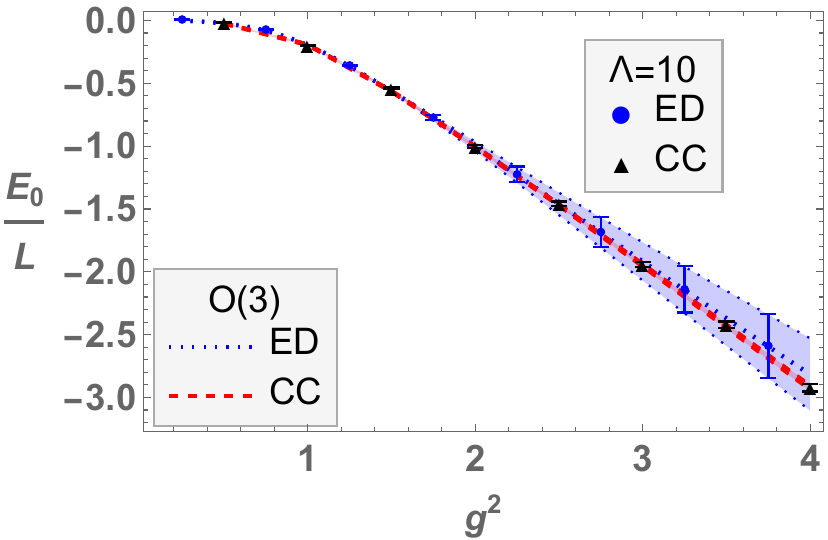}}
    \caption{Ground state energy $E_0/L$ for $L=2,3$ \textit{vs} $g^2$, computed with ED and the CC method, for $\Lambda=1,10$. ED is taken at $l_\text{max.}=3$, with error bars given by difference from $l_\text{max.}=2$ result. CC error bars for $L=3$ is statistical uncertainty from Monte Carlo integration.}
    \label{fig:lmda1_gd}
\end{figure}

\begin{figure}
    \centering
    \subfigure[$L=2$]{\includegraphics[scale=0.5]{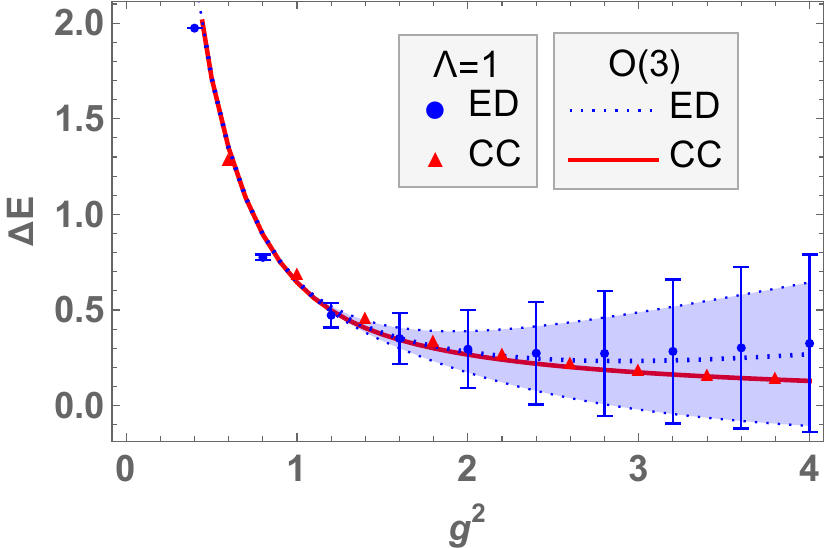}}
    \subfigure[$L=3$]{\includegraphics[scale=0.5]{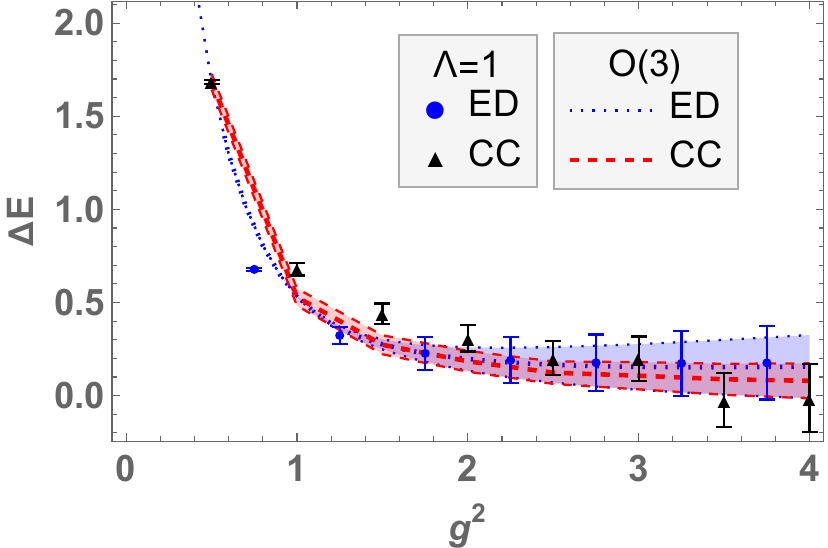}}
    \subfigure[$L=2$]{\includegraphics[scale=0.5]{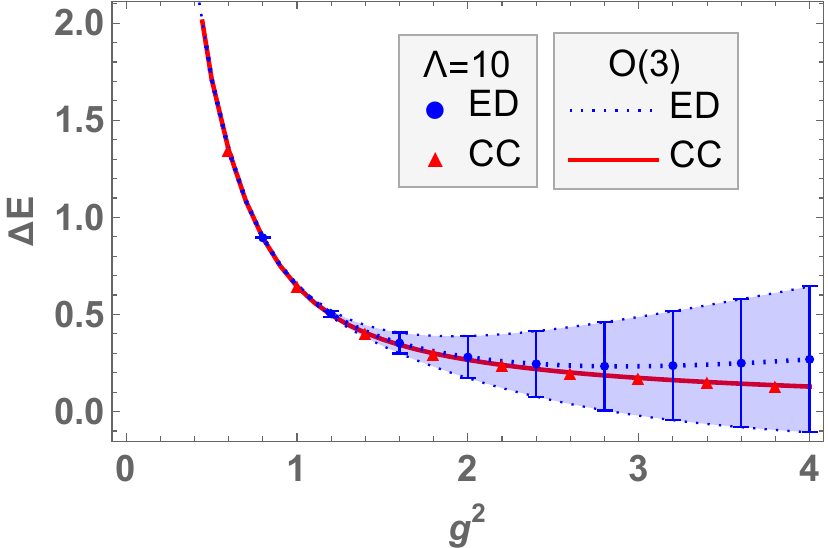}}
    \subfigure[$L=3$]{\includegraphics[scale=0.5]{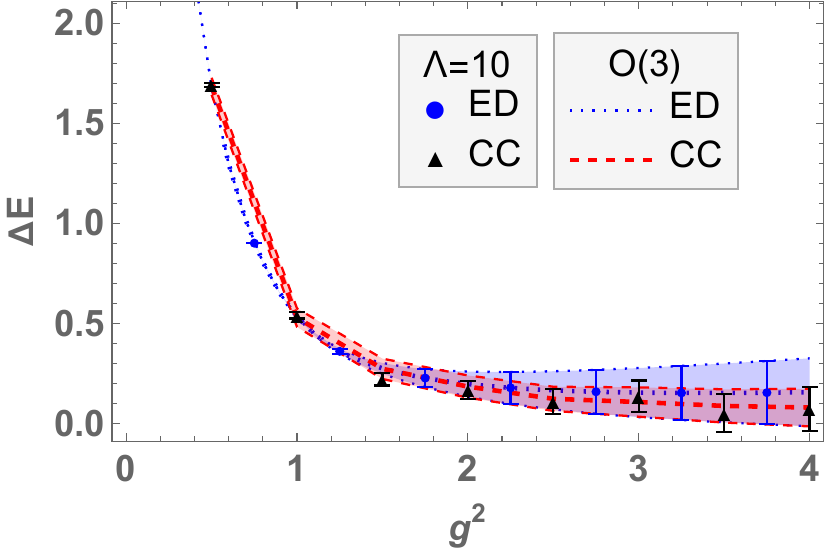}}
    \caption{Mass gap $\Delta E$ for $L=2,3$ \textit{vs} $g^2$, computed with ED and the CC method, for $\Lambda=1,10$. ED is taken at $l_\text{max.}=3$, with error bars given by difference from $l_\text{max.}=2$ result. CC error bars for $L=3$ is statistical uncertainty from Monte Carlo integration.}
    \label{fig:lmda1_gap}
\end{figure}

These states can be used as a basis for the matrix elements of the Hamiltonian \eqref{eq:og_H} from which we can calculate the energy levels.
To achieve a better understanding of the energy levels, we construct a variational Ansatz \cite{Peruzzo2014} by extending the CC method used for the $\text{O(3)}$ model in Section~\ref{sec:descr}. Analogous to (\ref{eq:CCOmega0}), we start with a reference state $\ket{\Omega(\Lambda)}$ defined as a tensor product of the states $\ket{l(x)=0,m(x)=0;\Lambda}$ (Eq.\ \eqref{eq:basis})
defined at each lattice site $x$,
\be \label{eq:Omega_def}\ket{\Omega(\Lambda)} \equiv \bigotimes_{x=0}^{L-1} \ket{l(x)=0,m(x)=0;\Lambda}\, . \ee
This is the weak coupling vacuum state which corresponds to the vanishing cluster operator $\hat{T} = 0$. We define $\hat{T}$ in terms of the potential term in the Hamiltonian and adopt the CC Ansatz:
\be \label{eq:CC_with_norms}\ket{\text{CC} (\Lambda)} \propto e^{-\frac{\alpha}{2L}\sum_x\left(\bm{\phi}(x) -\bm{\phi} (x+1)\right)^2} \ket{\Omega(\Lambda)}\ , \ee
which reduces to the CC Ansatz for the $\text{O(3)}$ model \eqref{eq:CC_0} in the limit $\Lambda\to\infty$. 

A similar approach based on a CC Ansatz which is a modified version of \eqref{eq:CC_with_norms} can be used to
estimate excited state energies. Notice that
\be \ket{l=1,m;\Lambda} \propto Y_{1m} (\bm{n}) \ket{l=0,m=0;\Lambda} \ee 
which is a direct consequence of Eq.\ \eqref{eq:basis}. Concentrating on $m=0$ (the other cases can be treated similarly), we define
\be\label{eq:420} \ket{\Omega_{1} (x,\Lambda)} \propto \phi_3(x) \ket{\Omega (\Lambda)}\,, \ee
and for the first excited state, we apply the following CC Ansatz
\be \label{eq:fe_cc} \ket{\text{CC}_{1} (\Lambda)} \propto e^{-\frac{\alpha}{2L} \sum_x \left(\bm{\phi}(x)-\bm{\phi}(x+1)\right)^2} \sum_x \ket{\Omega_{1}(x,\Lambda)} \,. \ee
This Ansatz reduces to the one for the $\text{O(3)}$ model, see Eq.\ \eqref{eq:fe_cc0}, in the limit $\Lambda\to\infty$.
The state $\ket{\Omega_{1}(x,\Lambda)}$, given by \eqref{eq:420},
is the weak-coupling eigenstate with site $x$ in the state $\ket{l(x)=1,m(x) =0;\Lambda}$, and all others in the $\ket{ l (x')=0,m(x')=0; \Lambda}$ state ($x'\ne x$). The degeneracy persists even at finite coupling, and so it should not matter which value of $m$ we choose at $x$. 

\begin{figure}[ht!]
    \centering
    \subfigure[$g^2=4$]{\includegraphics[scale=0.5]{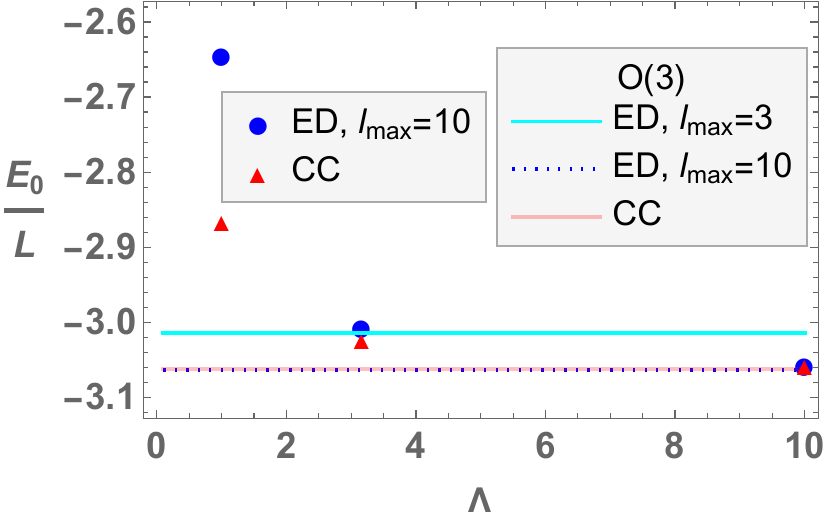}}
    \subfigure[$g^2=4$]{\includegraphics[scale=0.5]{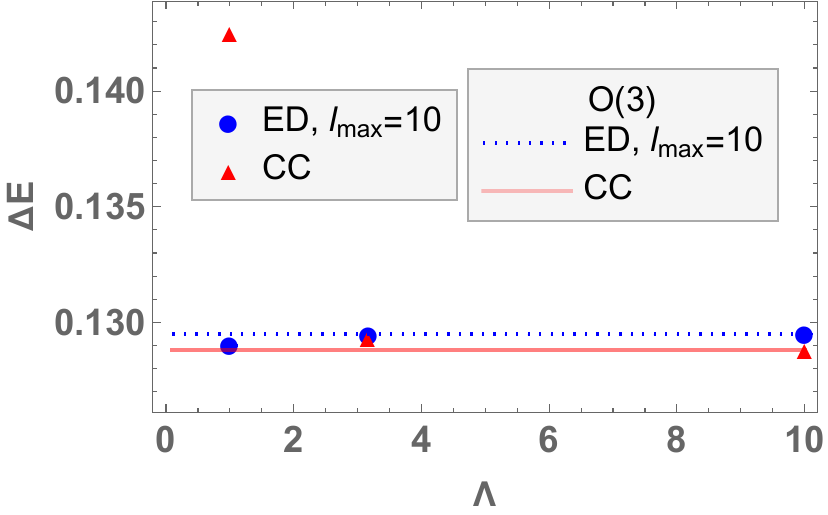}}
    \subfigure[$g^2=10$]{\includegraphics[scale=0.5]{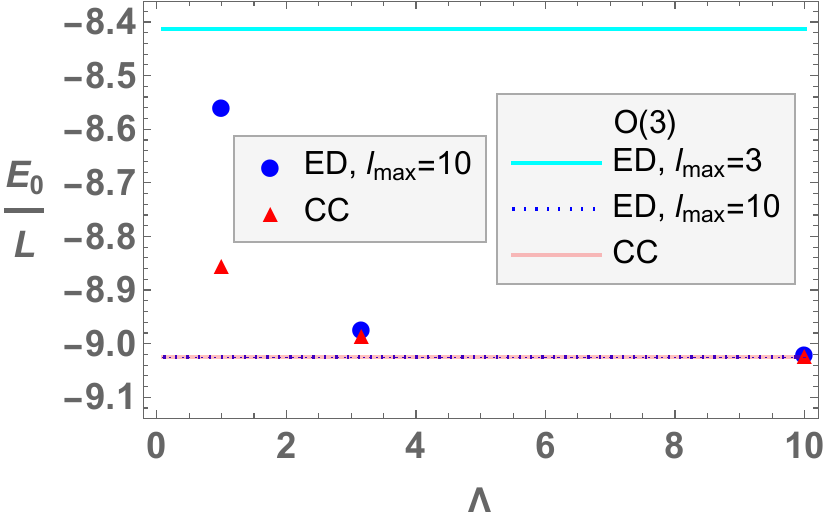}}
    \subfigure[$g^2=10$]{\includegraphics[scale=0.5]{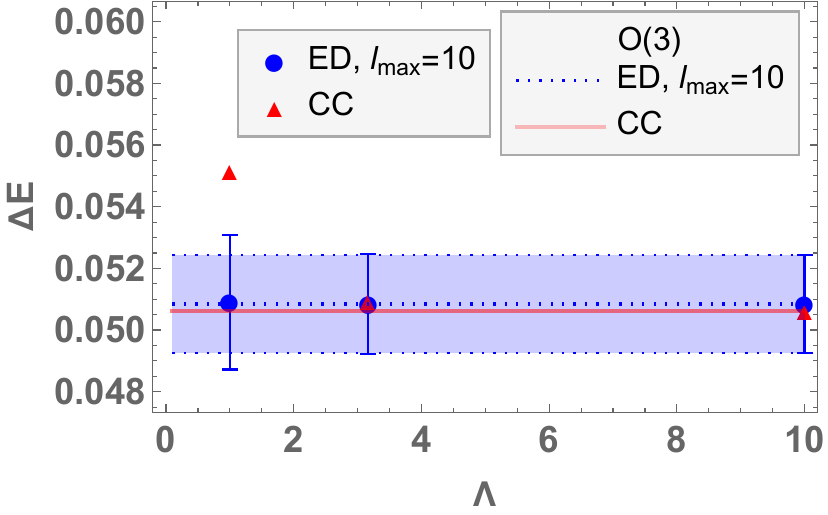}}
    \caption{Ground state energy $\frac{E_0}{L}$ and mass gap $\Delta E$ for $L=2$ sites, for $g^2=4$ (upper panels) and $g^2=10$ (lower panels) with $l_\text{max.}=10$. The displayed results include the $\text{O(3)}$ limit as well as three values for $\Lambda$: $1$, $3.2$ and $10$. Error bars for ED results are computed by finding the deviation from the $l_\text{max.}=9$ results.}
    \label{fig:extra_vs_no_extra}
\end{figure}

\begin{figure}
    \centering
    \subfigure[]{\includegraphics[scale=0.55]{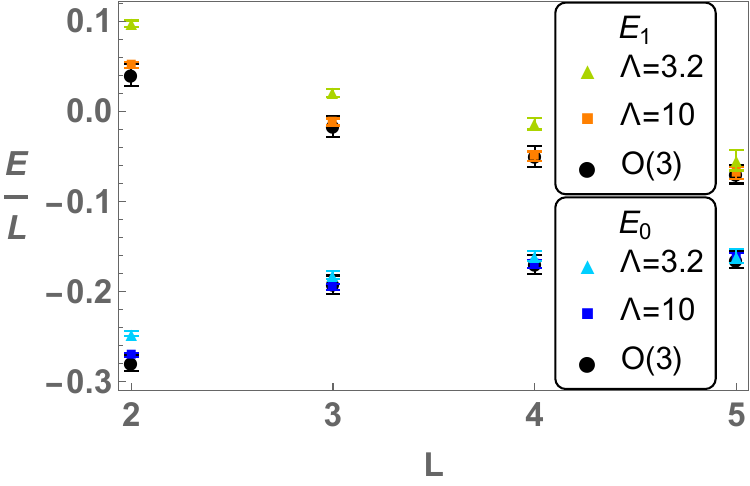}}
    \subfigure[]{\includegraphics[scale=0.55]{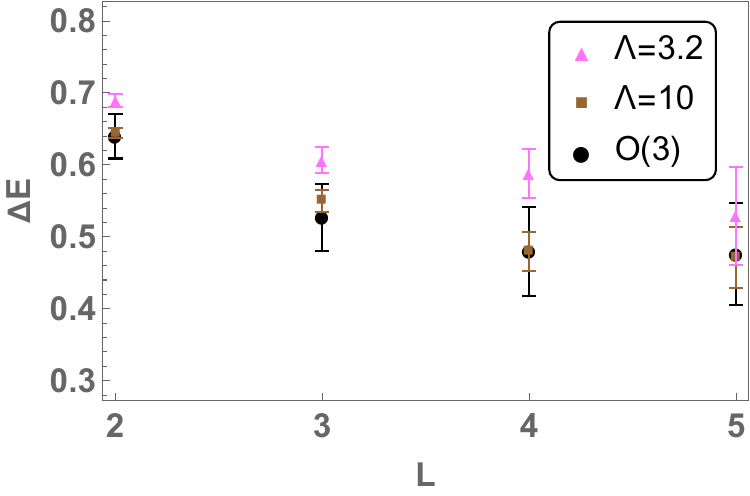}}
    \subfigure[]{\includegraphics[scale=0.55]{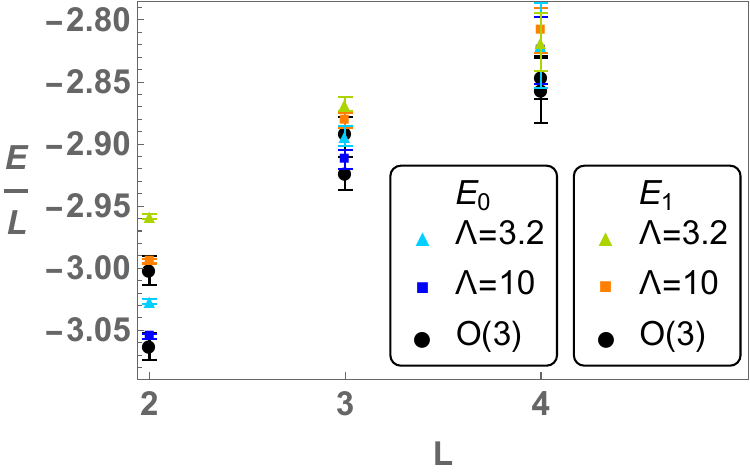}}
    \subfigure[]{\includegraphics[scale=0.55]{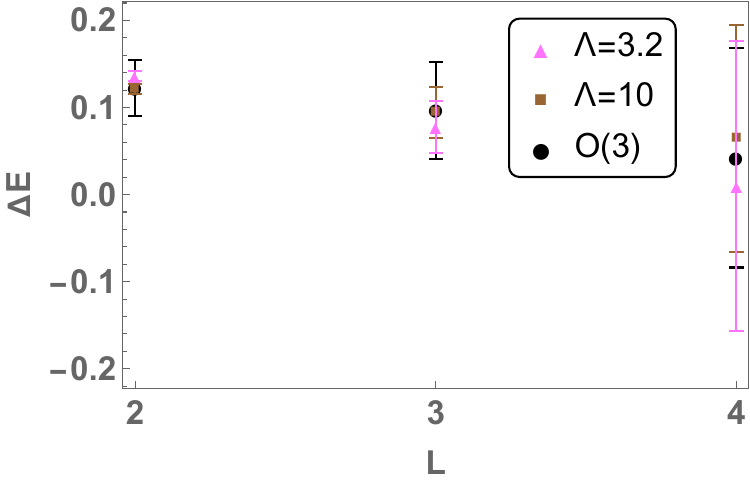}}
    \caption{(a) Ground and first excited state energies using the CC Ansatz for the $\text{O(3)}$ model, and for the qumode formulation for two values of $\Lambda$, with $g^2=1$ and up to $L=5$ sites. We used Monte Carlo integration with 500,000 sample points. (b) Mass gaps $\Delta E=E_1-E_0$ for $g^2=1$ as a function of $L$. (c) and (d): same except for $g^2=4$ and 5,000,000 sample points are used for the Monte Carlo integration.}
    \label{fig:sphere_E_vs_L}
\end{figure}

Next, we use the Hamiltonian given in \eqref{eq:og_H} with the local basis of states given in \eqref{eq:basis} to compute the ground state energy for different values of the cutoff $\Lambda$, once again comparing exact diagonalization with the coupled-cluster method. The results are shown in Fig.~\ref{fig:lmda1_gd}, 
which displays the ground state energy density for $L=2,3$, and $l_\text{max.}=3$ for ED. Here we consider two values of $\Lambda$, $1$ and $10$, and display the energies up to $g^2=4$. Fig.~\ref{fig:lmda1_gap} displays mass gaps for $L=2$ and $L=3$. 

To obtain the CC energies, we compute expectation values with numerical integration, with Monte Carlo integration being used for $L=3$ sites. In the latter, we first perform quasi-Monte Carlo integration \cite{Niederreiter1978} to estimate the optimal value of $\alpha$. Due to the deterministic nature of this method, we avoid noise in the minimization process. The energy expectation is then re-evaluated at the obtained value of $\alpha$ by sampling from a distribution whose radial component approximates the true radial wave function. For large values of $\Lambda$, a Gaussian distribution with mean $g$ and width $\sim \frac{1}{\Lambda}$ is sufficient, see Figs.~\ref{fig:lmda1_gd} and \ref{fig:lmda1_gap}(d). For smaller values of $\Lambda$, the radial wave function can not be accurately approximated by a Gaussian. In this case, it is best to use the full radial wave function at $\alpha=0$, which is the radial part of the integrand in Eq.~\eqref{eq:basis}, see Figs.~\ref{fig:lmda1_gd} and \ref{fig:lmda1_gap}(b). This allows us to obtain an accurate value of the energy with a lowered uncertainty compared with the quasi-Monte Carlo method. 

In Fig.~\ref{fig:extra_vs_no_extra}, we investigate the effectiveness of the CC Ansatz at large $g^2$. To this end, it is necessary to use a significantly higher value of $l_\text{max.}$ to provide an ED calculation to compare against. In this Figure, we set $l_\text{max.}=10$ and compute the ground state energies and mass gaps for two values of $g^2$ and three values of $\Lambda$. We not only find agreement between the two methods as $\Lambda$ becomes large, but we also find that the results of the CC calculation outperform the ED results for the $\text{O(3)}$ model with $l_\text{max.}=3$, as indicated by the light-blue lines in the two panels on the left-hand side.


For practical numerical simulations with matrices, anything beyond $L=4$, $l_\text{max.}=2$ where $H$ is of size $9^{4} \times 9^4$ presents a computational challenge.  Therefore, to obtain energy estimates for a larger number of sites, we proceed with numerical integration methods. In our CC results for $L=2$ displayed in Figures~\ref{fig:lmda1_gd} and \ref{fig:lmda1_gap}, the angular integrals were performed analytically, and a numerical integration over $r(x)$ was performed to complete the calculation of expectation values. For $L\ge 3$, however, we need to perform instead a Monte Carlo integration over all variables.

Numerical results for the energies of the ground and first excited state are displayed in Fig.~\ref{fig:sphere_E_vs_L} (top left panel) as a function of the number of lattice sites $L$, with the coupling constant set to $g^2=1$. The top right panel shows the mass gap $\Delta E$. It appears to converge near $\Delta E \simeq 0.48$ for a large number of lattice sites $L$. We observe the accumulation of numerical errors as the number of lattice sites is increased.  Analogously, the numerical results for $g^2=4$ are shown in the bottom two panels, with $10$ times as many sample points for the Monte Carlo integration. The gap at large $L$ is very small, as expected. The numerical errors are significantly larger compared to the $g^2=1$ case, making it difficult to determine the gap at more than four sites.

As with the $L=3$ results displayed in Figures~\ref{fig:lmda1_gd} and \ref{fig:lmda1_gap}, we approximate the optimal value of the CC parameter $\alpha$ with quasi-Monte Carlo integration, and then re-evaluate the energy expectation at that optimal value by sampling from a Gaussian distribution for the radial direction.

\section{Quantum simulation using continuous variables \label{sec:quantum_sim}}

In this section, we discuss the simulation of the $\text{O(3)}$ model using CV gates. We start with the construction of the CC Ansatz, which we then use for the quantum computation of energy levels. Next, we describe the relevant circuits for the time evolution within the CV approach and we present numerical results, which use the \textsc{Strawberry Fields} quantum simulator. More detailed discussions about CV gates can be found in Appendix~\ref{sec:CV_gates}.

\subsection{Coupled-Cluster Ansatz \label{section:CC}} 

\begin{figure}[ht]
    \centering
    \subfigure[$99\%$ fidelity]{\includegraphics[scale=0.4]{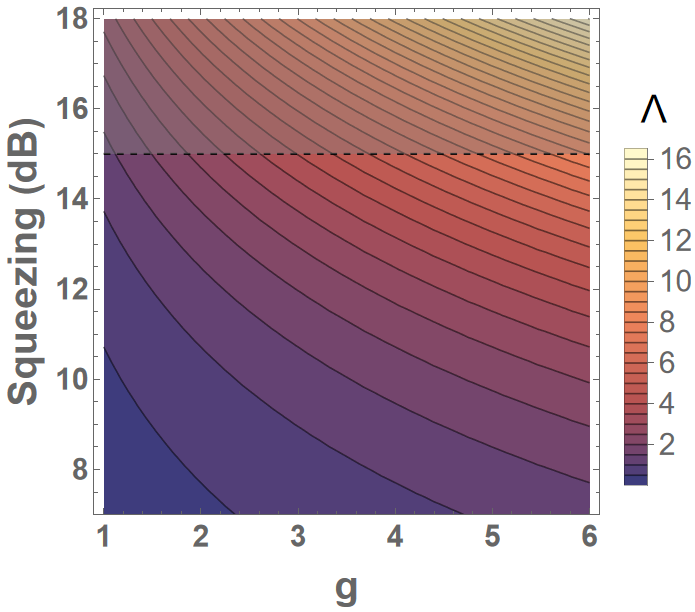}}
    \hspace{1em}
    \subfigure[$99.9\%$ fidelity]{\includegraphics[scale=0.4]{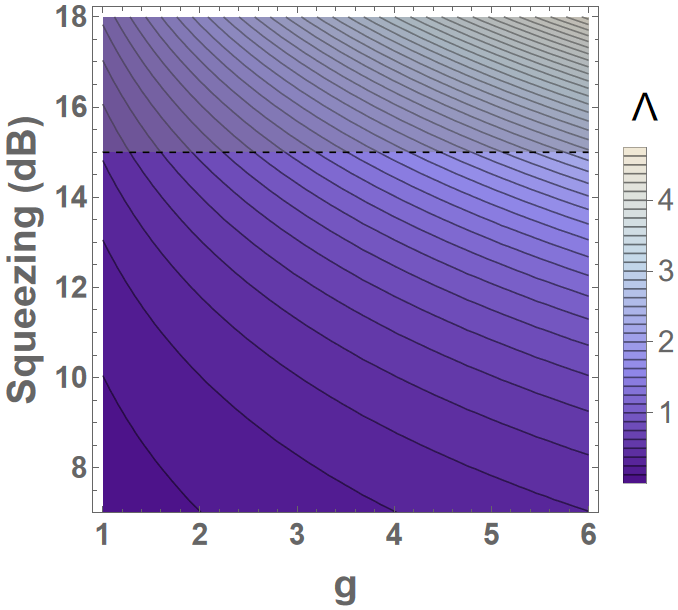}}
    \caption{Achievable values of $\Lambda$ in constructing the state $\ket{\omega(\Lambda)}$ as a function of coupling $g$ and squeezing in dB. Panels (a) \& (b) correspond to 99\% and 99.9\% cubic gate fidelity. Estimates are based on computed fidelities for gate parameter $s=0.3$ from Ref.\ \cite{Yanagimoto2020}, with the Kerr strength and gate time held to the values used for $s=0.3$. Displayed is the ideal (noiseless) case.}
    \label{fig:cubic_squeeze_fidelity}
\end{figure}

To engineer the CC Ansatz in \eqref{eq:CC_with_norms}, let us first concentrate on a single lattice site. By making use of three qumodes of quadratures $(q_a,p_a)$ ($a=1,2,3$), collectively denoted as $(\bm{q}, \bm{p})$, we initialize the system in the vacuum state 
\be \label{eq:triplet_vac} \ket{\bm{0}} \equiv \bigotimes_{a=1}^3 \ket{0}_a = \frac{1}{\pi^{3/4}}\int d^3 \bm{q}\, e^{-\frac{1}{2} \bm{q}^2} \ket{\bm{q}} \ee 
We attach an ancilla qumode of quadratures $(q_b, p_b)$, also initialized in the vacuum state 
\be 
\ket{0}_b = \frac{1}{\pi^{1/4}}\int d q_b\, e^{-\frac{1}{2} {q}_b^2} \ket{{q}_b}_b \,. 
\ee
We apply the product of two-mode entangling non-Gaussian unitaries
\be\label{eq:Uab} U_{ab} = e^{-i\frac{\Lambda}{\sqrt{2} g}  q_a^2 p_b } \,, \ee
followed by the application of a translation operator on the ancilla qumode
\be\label{eq:U_b} U_b = e^{{i}g\frac{\Lambda}{\sqrt2}\left(1+\frac{2}{\Lambda^2}\right)  \, p_b}\,. \ee
These operators shift the quadrature $q_b$ of the ancilla qumode. Altogether, we thus obtain
\be U_b \prod_{a=1}^3 U_{ab} \ket{\bm{0}} \ket{0}_b = \frac{1}{\pi}\int d^3 \bm{q}\, d q_b \, e^{-\frac{1}{2} \bm{q}^2}   e^{-\frac{1}{2} \left[{q}_b - \frac{\Lambda}{\sqrt{2}g} \left(\bm{q}^2 - \left(1+\frac{2}{\Lambda^2}\right)g^2\right)\right]^2} \ket{\bm{q}} \ket{{q}_b}_b \,.
\ee
Next, we measure the ancilla qumode projecting it onto the state $\ket{0}_b$. The resulting state, which we denote by $\ket{\omega(\Lambda)}$ only involves the three main qumodes. It is given by
\be\label{eq:278}  \ket{\omega(\Lambda)} \propto {}_b\!\bra{0} U_b \prod_{a=1}^3 U_{ab} \ket{\bm{0}} \ket{0}_b  \propto \int d^3 \bm{q}\, e^{- \frac{\Lambda^2}{8g^2} (\bm{q}^2 - g^2)^2} \ket{\bm{q}} \,. \ee
Evidently, we constructed
the state $ \ket{\omega (\Lambda)} = \ket{l=0,m=0,\Lambda}$ (Eq.\ \eqref{eq:basis}) which contributes to the state $\ket{\Omega(\Lambda)}$ (Eq.\ \eqref{eq:Omega_def}) used in the coupled-cluster Ansatz. The associated circuit in terms of CV gates is shown in Fig.~\ref{fig:4}.
\begin{figure}[t!]
    \centering
\[\Qcircuit @C=1.2em @R=1em { \lstick{\ket{0}_1} &\qw & \qw &\qw & \qw  &\qw & \qw& \qw & \link{1}{-1} &   \qw  &\qw &\qw & \qw & \link{1}{-1} & \qw \\  \lstick{\ket{0}_2} & \qw & \qw & \link{1}{-1} & \qw & \qw & \link{1}{-1} & \qw & \link{-1}{-1} & \link{1}{-1}  &\qw &\qw  & \link{1}{-1} & \link{-1}{-1} &\qw \\ \lstick{\ket{0}_3} & \multigate{1}{U_{3b}} &\qw & \link{-1}{-1} & \multigate{1}{U_{2b}} & \qw & \link{-1}{-1} & \qw & \qw  &\link{-1}{-1}& \multigate{1}{U_{1b}} &\qw & \link{-1}{-1} & \qw &\qw \\ \lstick{\ket{0}_b} & \ghost{U_{3b}}   & \qw  &\qw & \ghost{U_{2b}} & \qw & \qw & \qw & \qw   &\qw& \ghost{U_{1b}}  &\qw& \qw  &\gate{U_b} &\measureD{n=0}   } \]
    \caption{CV quantum circuit for the generation of the state  $\ket{\omega(\Lambda)}=\ket{l=0,m=0,\Lambda}$ (Eq.\ \eqref{eq:278}) relevant for the construction of the coupled-cluster Ansatz.}
    \label{fig:4}
\end{figure}
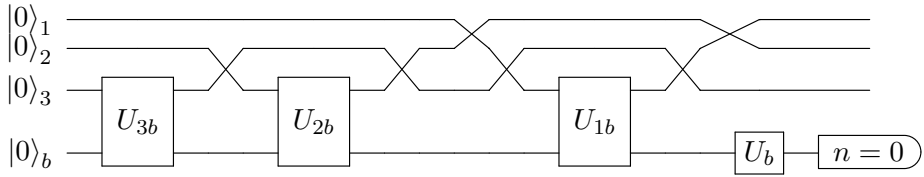
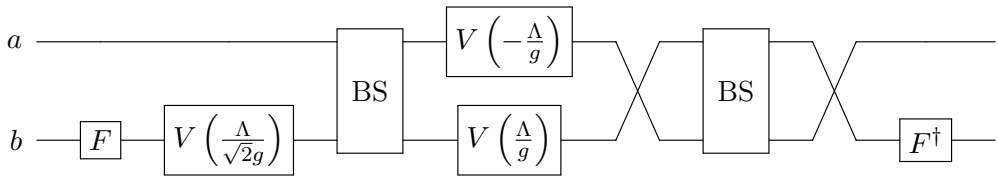
\begin{figure}[ht!]
    \centering
\[\Qcircuit @C=1.5em @R=1em {   \lstick{a} & \qw  &\qw   & \multigate{1}{\text{BS}} &\gate{V\left(-\frac{\Lambda}{g}\right)}&\qw&\link{1}{-1} & \multigate{1}{\text{BS}}   &\qw &\link{1}{-1} &\qw&\qw\\ \lstick{b} & \gate{F}   &\gate{V\left(\frac{\Lambda}{\sqrt{2} g}\right)}  &\ghost{\text{BS}} & \gate{V\left(\frac{\Lambda}{g}\right)}&\qw&\link{-1}{-1} &\ghost{\text{BS}} &\qw &\link{-1}{-1} &\gate{F^\dagger} &\qw } \]
    \caption{CV quantum circuit (see Eq.\ \eqref{eq:prepare_U2}) implementing the non-Gaussian unitary operator $U_{ab}$ given in \eqref{eq:Uab}.}
    \label{fig:4a}
\end{figure}
The non-Gaussian unitaries $U_{ab}$  in~\eqref{eq:Uab} can be expressed in terms of cubic phase gates acting on the ancilla qumode. Using the relation 
\begin{equation}
    6 q_a^2 p_b  = (q_a+ p_b)^3-(q_a-p_b)^3 - 2p_b^3 \,,
\end{equation}
we obtain the following gate decomposition of $U_{ab}$:
\be\label{eq:prepare_U2} U_{ab} =  F_b^\dagger \cdot BS_{ba} \cdot V_a \left(-\frac{\Lambda}{g}\right) \cdot  V_b \left( \frac{\Lambda}{g}\right) \cdot BS_{ab} \cdot V_b \left(\frac{\Lambda}{\sqrt{2}g}\right) \cdot F_b\ . \ee
where $F$ is the Fourier transform operator (see \eqref{eq:aF}), $BS_{ab}$ implements a 50:50 beam splitter ($\theta = \frac{\pi}{4}$ in the definition \eqref{eq:BS_def}), and we introduced the cubic phase gate
\be V(s) = e^{i\frac{s}{3} q^3} \ . \ee 
The quantum circuit for this unitary is shown in Fig.\ \ref{fig:4a}. Ref.~\cite{Yanagimoto2020} demonstrates that the cubic phase gate can be constructed from Kerr operations instead of having to use a measurement-based method. 
It was shown that the cubic gate parameter is
$s=\frac{3}{\sqrt2} \chi\tau\alpha \lambda^3$,
where $\alpha$ is a displacement parameter and $\lambda$ corresponds to a squeeze parameter $r=\ln{\lambda}$. $\chi$ characterizes the Kerr non-linearity and $\tau$ the Kerr gate time. The parameter $\alpha$ was adjusted to reduce algorithmic error and becomes $\alpha\sim \lambda^3$ in the noiseless case. In Fig.\ \ref{fig:cubic_squeeze_fidelity}, we estimate achievable values of the cutoff $\Lambda$ as a function of squeezing using this method, in the absence of losses. The squeezing is calculated by first considering the required amount to implement the V gate with gate parameter $s=0.3$ at approximately 99\% and 99.9\%, respectively, the approximate values of which are displayed in Ref.~\cite{Yanagimoto2020}. We then added the additional number of decibels of squeezing to increase the gate parameter to $s=\frac{\Lambda}{g}$. 

Next, we consider two lattice sites each occupied by qumode triplets of quadratures $({q}_a ,{p}_a)$ and $({q}_{a'}, {p}_{a'})$, respectively with $a, a'=1,2,3$. As  before, we collectively denote them by $(\bm{q}, \bm{p})$ and $(\bm{q}', \bm{p}')$.  Analogous to the procedure discussed above, we start by engineering the state
\be \ket{\omega (\Lambda)}_a \otimes \ket{\omega (\Lambda)}_{a'} = \ket{l=0,m=0;\Lambda}_a\otimes \ket{l' =0,m'=0; \Lambda}_{a'} \,, \ee
which is given in terms of a direct tensor product of the states constructed for each lattice site, see (\ref{eq:278}). We proceed by attaching a triplet of ancilla qumodes of quadratures $(\bm{q}_c,\bm{p}_c)$ initialized in the vacuum state, and apply the Gaussian unitary (a product of three CX gates \eqref{eq:aCX})
\be W_{ac} = e^{-i\sqrt{\frac{2\alpha}{L}} \bm{q}\cdot \bm{p}_c} \ee 
followed by the analogous unitary $W_{a'c}^\dagger$. These two operators shift the quadratures of the ancilla qumodes. We obtain the state

\bea W_{a'c}^\dagger W_{ac} \ket{\omega (\Lambda)}_a \otimes \ket{\omega (\Lambda)}_{a'} &=& \int d^3 \bm{q}\,  d^3 \bm{q}'\,  d^3 \bm{q}_c\, e^{-\frac{1}{2} \bm{q}_c^2}  e^{- \frac{\Lambda^2}{8g^2} (\bm{q}^2 - g^2)^2}   e^{- \frac{\Lambda^2}{8g^2} ({\bm{q}'}^2 - g^2)^2} \nonumber\\ && \times  \ket{\bm{q}}   \ket{\bm{q}'} \ket{\bm{q}_c +\sqrt{\frac{2\alpha}{L}} (\bm{q} -\bm{q}')} \,. \eea

\begin{figure}[t!]
    \centering
\[\Qcircuit @C=1.8em @R=1em {   \lstick{\ket{\omega(\Lambda)}_{a'}} & \qw  &\qw & \link{1}{-1}  & \qw  & \qw & \link{1}{-1}    &\qw  \\ \lstick{\ket{\omega(\Lambda)}_{a}} & \multigate{1}{W_{ac}}   &\qw & \link{-1}{-1} &\multigate{1}{W_{a'c}^\dagger}  & \qw  & \link{-1}{-1} &\qw  \\
\lstick{\ket{\bm{0}}_c} &\ghost{W_{ac}}& \qw &\qw & \ghost{W_{a'c}^\dagger} & \qw &\qw & \measureD{n=0} }  \]
    \caption{CV quantum circuit for constructing the CC state for two lattice sites in \eqref{eq:state}. Each line represents three qumodes. } 
    \label{fig:4b}
\end{figure}
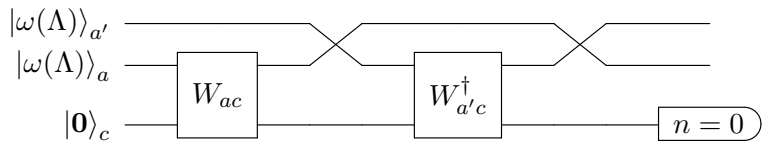

After measuring the ancilla qumodes and projecting them onto the vacuum state $\ket{\bm{0}}_c$, we obtain the state
\be\label{eq:state} e^{-\frac{\alpha}{2L} (\bm{q} - \bm{q}')^2} \ket{\omega (\Lambda)}_a \otimes \ket{\omega (\Lambda)}_{a'} \,. \ee 
After identifying $\bm{\phi} (x) = \bm{q}$, $\bm{\phi} (x+1) = \bm{q}'$, we have thus engineered the desired state $\ket{{\rm CC}(\Lambda)}$ for two lattice sites, see \eqref{eq:CC_with_norms}.  The circuit to construct this state is shown in Fig.~\ref{fig:4b}. This can be generalized to an arbitrary number of lattice sites by repeating the above procedure for each pair of adjacent sites $(x,x+1)$ of the 1-dimensional lattice.

\begin{figure}[ht]
    \centering
    \subfigure[$g^2=1$]{\includegraphics[scale=0.5]{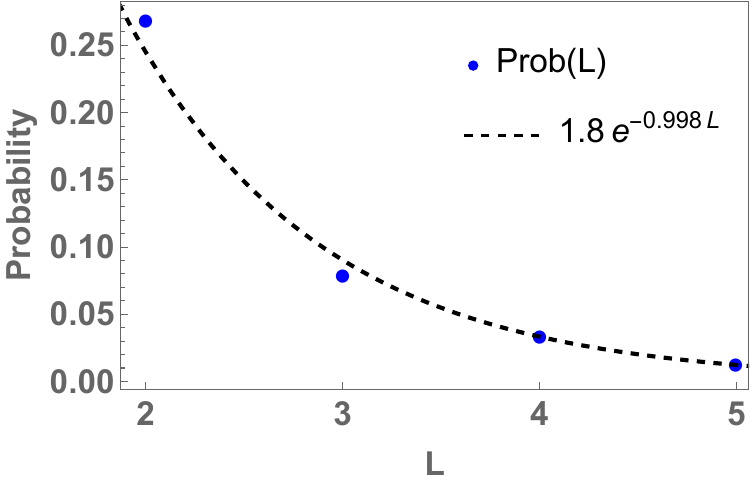}}
    \hspace{2em}
    \subfigure[$g^2=4$]{\includegraphics[scale=0.5]{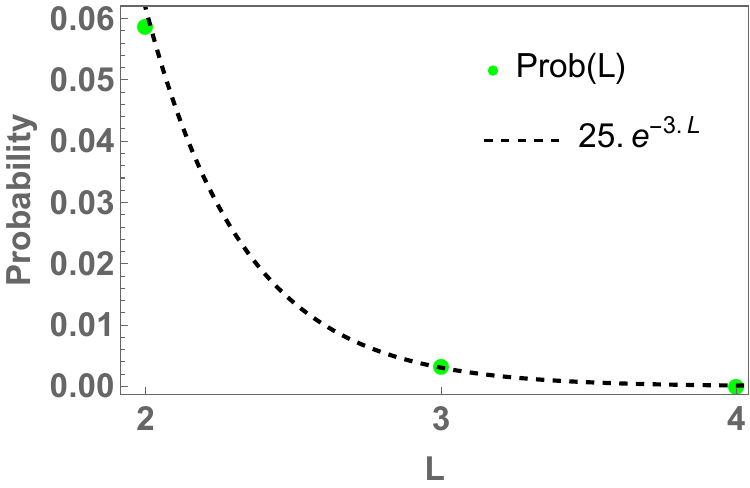}}
    \caption{Success probability of obtaining the CC state (at the optimal value of $\alpha$) as a function of lattice size $L$, given an ensemble of resource states $\ket{\omega(\Lambda)}$ (prepared using Fig.\ \ref{fig:4}). The points are fitted to an exponential function of $L$.}
    \label{fig:prob_of_CC}
\end{figure}

As our scheme is not fully unitary and requires ancilla measurements, it is important to analyze the success rate of our algorithm. Clearly, the assignment of separate ancillae to each spatial site suggests that the success rate decreases with lattice size. Notice that the circuits of Figs.\ \ref{fig:4} and \ref{fig:4b} both require ancilla measurements on each site. Fig.\ \ref{fig:4} corresponds to the generation of resource states for the calculation. Since they are site-local, this generation can be done in parallel and offline (for $g=1$ and $\Lambda=3.2$ the success rate to generate $\ket{\omega(\Lambda)}$ is $\sim 0.4$, so that it would only take $\sim 2\text{-}3$ runs to obtain the state, with the number of runs increasing with $g$ and $\Lambda$). Therefore, we compute the success rate of Fig.\ \ref{fig:4b} assuming that we have access to the resource states $\ket{\omega(\Lambda)}$. These rates are shown in Fig.\ \ref{fig:prob_of_CC}. Fitting to an exponential in the lattice size $L$, the rate of decay appears to be $g^2-1$.

Given the exponential decrease in the success probabilities with lattice size, it may be worth identifying a unitary coupled-cluster (UCC) Ansatz through which we would not need to perform so many photon number measurements for large $L$. This is a subject for future work. 

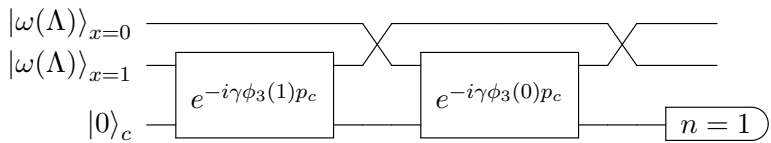
\begin{figure}[t!]
    \centering
\[ \ \ \ \ \ \ \ \ \  \ \Qcircuit @C=1.em @R=1em {   \lstick{\ket{\omega(\Lambda)}_{x=0}} & \qw  &\qw & \link{1}{-1}  & \qw  & \qw & \link{1}{-1}    &\qw  \\ \lstick{\ket{\omega(\Lambda)}_{x=1}} & \multigate{1}{e^{-i\gamma \phi_3(1)p_c}}   &\qw & \link{-1}{-1} &\multigate{1}{e^{-i\gamma \phi_3(0)p_c}}  & \qw  & \link{-1}{-1} &\qw  \\
\lstick{\ket{{0}}_c} &\ghost{e^{-i\gamma \phi_3(1)p_c}}& \qw &\qw & \ghost{e^{-i\gamma \phi_3(0)p_c}} & \qw &\qw & \measureD{n=1} }  \]
    \caption{CV quantum circuit relevant for constructing the CC ansatz for the first excited state, see \eqref{eq:fe_cc}, with $L=2$ lattice sites.}
    \label{fig:4c}
\end{figure}

Next, we consider the construction of the first excited state, see \eqref{eq:fe_cc} above. The construction of this state is a direct extension of the procedure for the ground state given in \eqref{eq:CC_with_norms}, which we have already engineered.
After obtaining the state $\ket{\Omega (\Lambda)}$ as a tensor product of the states $\ket{\omega(\Lambda)}$ (Eq.\ \eqref{eq:278}) at each lattice site, we add an ancilla of quadratures $(q_c,p_c)$ prepared in the vacuum state $\ket{0}_c$. We then apply the following string of Gaussian unitaries (CX gates)
\be \prod_x e^{-i\gamma \phi_3(x)\otimes p_c}\, ,\ \gamma\ll 1\, , \ee
and measure the added ancilla in the photon number basis projecting it onto the single-photon state. It is important that $\gamma$ be chosen to be sufficiently small to avoid introducing an unwanted dependence of the resulting wave function on $\phi_3 (x)$. We obtain the following uniform superposition of states
\be\label{eq:1stCC} 
{}_c\bra{1} \prod_x e^{-i\gamma \phi_3(x)\otimes p_c} \ket{\Omega (\Lambda)}\ket{0}_c \propto  \sum_x \phi_3 (x) \ket{\Omega (\Lambda)} \,, \ee
on which the CC Ansatz for the first-excited state \eqref{eq:fe_cc} is based. The quantum circuit for $L=2$ lattice sites is shown in Fig.~\ref{fig:4c}. After obtaining the state in \eqref{eq:1stCC}, we still need to apply the circuit shown in Fig.~\ref{fig:4b} to realize the full CC Ansatz analogous to the ground state described above.

\subsection{\label{subsec:protocol_E_TE}Quantum computation of energy levels}

Having constructed the CC Ansatz, we now introduce the procedure to obtain the ground state energy of the Hamiltonian given 
in \eqref{eq:og_H} using quantum resources. We compute energies using the Variational Quantum Eigensolver (VQE) algorithm \cite{Peruzzo2014,Bharti2022}, which has seen use for simulation in various contexts \cite{Hempel2018,Xu2020,Dumitrescu:2018njn,Yeter2019}. In our case, the parameterized CC Ansatz becomes the input for the energy functional $\bra{\text{CC}(\alpha)}H\ket{\text{CC}(\alpha)}$, where $H$ is the Hamiltonian \eqref{eq:og_H}. The energy functional is computed on a quantum computer by first choosing a computational basis, writing $H$ as a sum of terms each of which can be diagonalized by a unitary $U$ in that basis, and then applying $U$ to the variational Ansatz to obtain a set of modified circuits for quantum computation. We will discuss these modifications in a CV quantum computing context below (where the computational basis is the photon number basis). Each circuit is then measured (sampled) in the chosen basis and the resulting contribution to the total functional $\langle H\rangle$ is given by the sample mean of these measurements.  The output of the functional is minimized by optimizing over our parameters, just $\alpha$ in our case, using a classical optimizer. In our numerical simulations, we utilise Nelder-Mead optimization \cite{Nelder1965} which is a gradient-free technique.

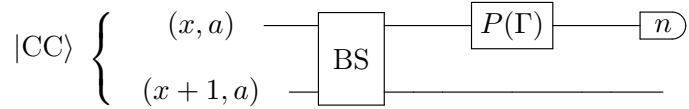
\begin{figure}[t!]
    \centering
\[ \ \ \ \ \ \ \ \ \  \ \Qcircuit @C=1.em @R=1em {    \lstick{} & \push{(x,a)\ \ \ } & \multigate{1}{\text{BS}}   &\qw & \qw &\gate{P(\Gamma)}  & \qw  & \qw &\measureD{n}  \\
\lstick{} & \push{(x+1,a)\ \ \ } &\ghost{\text{BS}}& \qw &\qw & \qw & \qw &\qw & \qw  \inputgroupv{1}{2}{.8em}{.8em}{\ket{\text{CC}} \ \ \ \ } }  \]
    \caption{Quantum circuit for the calculation of the expectation value of the interaction term in the Hamiltonian \eqref{eq:og_H} involving $\phi_a (x)$ and $\phi_a (x+1)$. }
    \label{fig:4aa}
\end{figure}

The first term in the expectation value $\langle H\rangle$ we consider is that of the interaction term
$\frac{1}{2} \sum_x \left(\bm{\phi}(x) - \bm{\phi}(x+1)\right)^2$.
After engineering the CC Ansatz, see \eqref{eq:CC_with_norms}, we fix $x$ and create the state
\be \ket{\Psi(x,\Gamma)}\equiv \prod_a P_{x,a} (\Gamma) \cdot BS_{x,x+1}^a \ket{\text{CC}} \ee
by acting on the CC Ansatz first with a series of 50:50 beam splitters with neighboring lattice sites as input ports followed by quadratic phase gates $P_{x,a}(\Gamma) = e^{i\Gamma {\phi}_a^2(x)/2}$, where $\Gamma$ is a real parameter (see Eq.\ \eqref{eq:aP}). Note that 
the quadratic phase gate can be decomposed in terms of a single rotation and single-mode squeezer gate. We then compute
\bea\label{eq:417} \frac{1}{\Gamma^2}\sum_x \sum_a \Big[ \bra{\Psi(x,\Gamma)} N_{x,a} \ket{\Psi(x,\Gamma)} +&& \bra{\Psi(x,-\Gamma)} N_{x,a} \ket{\Psi(x,-\Gamma)}\nonumber\\-&& 2\bra{\Psi(x,0)} N_{x,a} \ket{\Psi(x,0)} \Big] \,, \eea
where $N_{x,a}$ is the number operator for the qumode labeled by $x,a$, which can be written as $N_{x,a} = \frac{1}{2} ( \pi_a^2 (x) + \phi_a^2 (x))$ with $a=1,2,3$. To see why this gives us the expectation value of the interaction term of the Hamiltonian in \eqref{eq:og_H}, note that for a single qumode of quadratures $(q,p)$, we have
\be P^\dagger (\Gamma) N P(\Gamma) = \frac{1}{2}\left(q^2+\left(p+\Gamma q\right)^2\right)\,, \ee
where $N = \frac{1}{2} (p^2 + q^2)$ and $P(\Gamma)$ is the quadratic phase gate as introduced above. We deduce the following parameter shift rule
\be \frac{1}{\Gamma^2} \left[ P^\dagger (\Gamma) N P(\Gamma) + P^\dagger (-\Gamma) N P(-\Gamma) - 2N \right] = q^2 \,.\ee
This yields the expression involving expectation values given in \eqref{eq:417}
by taking $q=\phi_a(x)$, and noting that applying the 50:50 beam splitters results in
\be\phi_a(x)\to \frac{1}{\sqrt2}\left(\phi_a(x)-\phi_a(x+1)\right)\ee
Therefore,
we obtain the desired expression for each one of the interaction terms in the Hamiltonian \eqref{eq:og_H}. A quantum circuit for the above calculation is shown in Fig.~\ref{fig:4aa} for qumodes $\phi_a (x)$ and $\phi_a (x+1)$. 

For practical purposes, when computing expectation values, it is advantageous to add an ancilla qumode of quadratures $(q_c,p_c)$ prepared in the vacuum state and to use a $CX$ gate instead of a $P$ gate. That is, we make the replacement
$e^{i\Gamma \phi_a^2(x)/2}\to e^{-i\Gamma p_c\otimes\phi_a(x)}$.
This reduces numerical errors at small values of $\Gamma$. It is necessary for $\Gamma$ to be small not only to lower truncation errors on a classical simulator, but also to keep additional squeezing minimal. 
Instead of a parameter shift rule, we need to choose only one value of $\Gamma$ and compute
\be\label{eq:CX_not_P_1}\frac{2}{\Gamma^2}\langle\Psi(\Gamma)\vert N\vert\Psi(\Gamma)\rangle \,.\ee 
This gives the desired result due to the relation
\be\label{eq:CX_not_P_2} CX \left(-\Gamma\right)\cdot N_c\cdot CX\left(\Gamma\right)=N_c+\Gamma q_c \phi_a(x)+\frac{\Gamma^2}{2}\phi_a^2(x)\, ,\ee
where the expectation values of $N_c$ and $q_c$ vanish for the vacuum state, which the ancilla is prepared in. The only remaining term is the one we need, thereby preventing errors derived from unphysical terms.
However, with only one ancilla this method requires us to compute 
$ \langle \left(\Delta\phi_a (x)\right)^2\rangle$
for each value of $x$ and $a$. 

Next, we are going to consider the kinetic energy term in the Hamiltonian \eqref{eq:og_H}, which is given by
\be \frac{1}{2g^2} \sum_x \bm{ L}^2(x) = \frac{1}{2g^2} \sum_x \left(\bm{\phi}(x) \times \bm{\pi} (x)\right)^2\,.\ee
To obtain its expectation value, we start by fixing $x$ and $a=3$, and consider $\langle L_3^2(x)\rangle$. The contributions of the other components can be treated similarly. We compute the expectation value of the square of the number operator for the $a=1,2$ qumodes at lattice site $x$ as 
\be \Delta N_{12} (x) \equiv \left(N_2(x) - N_1(x) \right)^2 \,,
\ee 
with respect to the state 
\be \label{eq:Lpsi_def} \ket{\Psi}\equiv BS_x^{12} \cdot F_{x,1} \ket{\text{CC}} \,. \ee
Here $F$ is again the Fourier transform operator $F = e^{i\frac{\pi}{4} (p^2 + q^2)}$ and $BS$ is a 50:50 beam splitter. After some algebra, we obtain the following relation
\be\label{eq:425} L_3^2 (x) = F_{x,1}^\dagger \cdot BS_x^{12} \cdot \Delta N_{12} (x) \cdot BS_x^{12} \cdot F_{x,1}  \,. \ee
hence
\be\label{eq:426}  \bra{\text{CC}} {L_3^2(x)} \ket{\text{CC}} = \bra{\Psi} \Delta N_{12} (x) \ket{\Psi} \ee
The expectation values of $L_{1,2}^2 (x)$ can be obtained by cyclic permutation of the indices.

\begin{figure}
    \centering
    \subfigure[$g^2=1,\alpha=0.83$]{\includegraphics[scale=0.45]{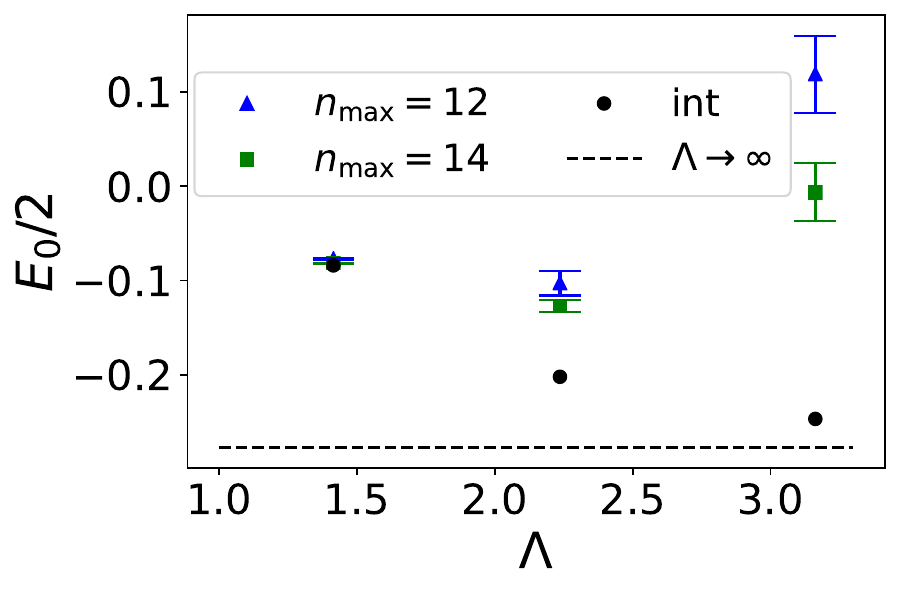}}
    \hspace{1em}
    \subfigure[$g^2=4,\alpha=1$]{\includegraphics[scale=0.45]{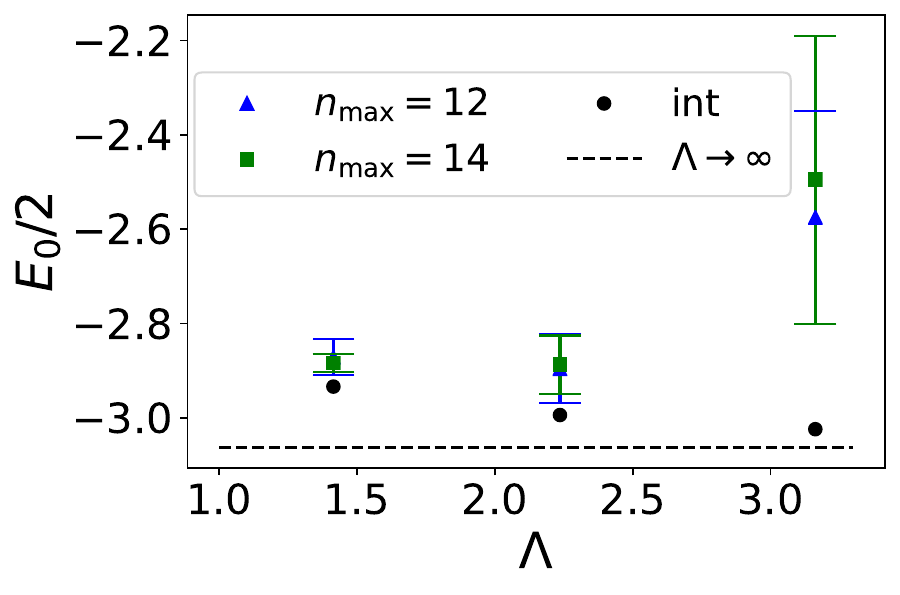}}
    \caption{\SF\ simulation results for the calculation of the ground state energy using the CC Ansatz near the minimizing value of $\alpha$ for (a) $g^2=1$ and (b) $g^2=4$ and $L=2$. The black points were obtained with numerical integration over the radial coordinates, and the dashed line obtained using Eq.\ \eqref{eq:Leq2_analytic}.}
    \label{fig:L2_Es_sf}
\end{figure}

\begin{figure}
    \centering
    \subfigure[$g^2=1,\alpha=0.83$]{\includegraphics[scale=0.45]{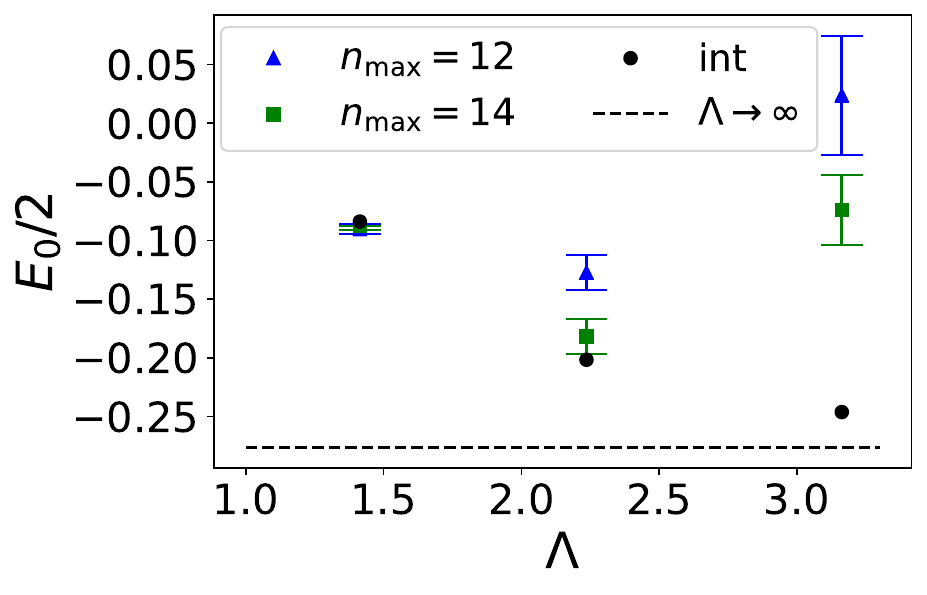}}
    \hspace{1em}
    \subfigure[$g^2=4,\alpha=1$]{\includegraphics[scale=0.45]{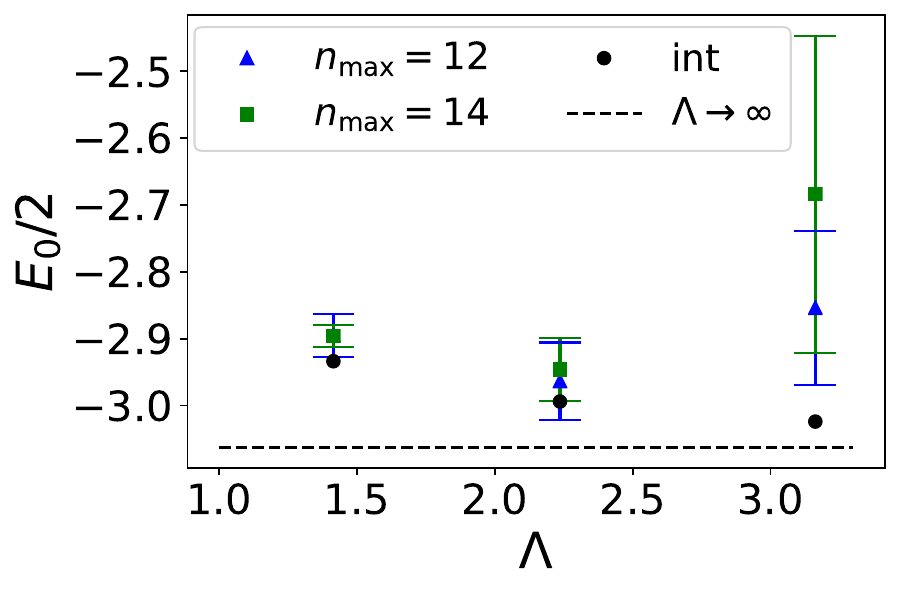}}
    \caption{\SF\ simulation results for the calculation of ground state coupled-cluster energies near the minimizing value of $\alpha$ for (a) $g^2=1$ and (b) $g^2=4$ and $L=2$. Here we use \eqref{eq:Lsq_method2} to compute the kinetic energy. The black points were obtained with numerical integration over the radial coordinates, and the dashed line was obtained using Eq.\ \eqref{eq:Leq2_analytic}.}
    \label{fig:L2_Es_sf_method2}
\end{figure}

Next, we present numerical results using the quantum simulator \SF\ \cite{Killoran_2019} for two values of the coupling, $g^2=1$ and $g^2=4$. For these simulations, we need to impose a Fock space cutoff, which we denote by $n_\text{max}$. The results for the ground state using the CC ansatz are shown in Fig.~\ref{fig:L2_Es_sf}. These were found by computing the energy expectation near the optimal value of $\alpha$ obtained from numerical integration. For the coupling term, we employed an ancilla and $CX$ gates instead of $P$ gates, see Eqs.~\eqref{eq:CX_not_P_1} and \eqref{eq:CX_not_P_2} above. The displayed error bars were obtained by computing the energy along each Cartesian direction. Since the CC ground state is invariant under global rotations, each result (multiplied by $3$) should be valid, and any variation reflects truncation errors. Even at $n_\text{max}=14$, which requires a Hilbert space dimension of $14^7$ ($3L$ physical qumodes plus an ancilla summing to 7), we see that there is very significant truncation error for $\Lambda=3.2$. Therefore, we conclude that it requires an immense amount of computational resources to simulate a CC state with a sufficiently large value of $\Lambda$ to estimate the $\text{O(3)}$ energy. The results at higher $n_\text{max}$ seem to be improved somewhat if we compute $\braket{\bm{L}^2}$ using Eq.\ \eqref{eq:426} but by breaking up the expectation value as
\be\label{eq:Lsq_method2} \bra{\text{CC}}L_3^2(x)\ket{\text{CC}}=2\left(\bra{\Psi}N_1^2(x)\ket{\Psi}+\bra{\Psi}N_2^2(x)\ket{\Psi}\right)-\bra{\text{CC}}\left(N_1+N_2\right)^2\ket{\text{CC}}, \ee
see Fig.~\ref{fig:L2_Es_sf_method2}. Here the last term is an expectation in the state $\ket{\text{CC}}$ rather than $\ket{\Psi}$ (Eq.~\eqref{eq:Lpsi_def}), to be compared with Fig.~\ref{fig:L2_Es_sf} which uses $\ket{\Psi}$ for all terms:
\be\bra{\text{CC}}L_3^2(x)\ket{\text{CC}}= \bra{\Psi}N_1^2(x)\ket{\Psi} + \bra{\Psi}N_2^2(x)\ket{\Psi} -2\bra{\Psi}N_1N_2\ket{\Psi} \ee
In any case, we note that the truncation error is in the number basis, unlike our ED results which were truncated in the spherical harmonic basis. 

\subsection{Time evolution circuits and photonic quantum simulator results}

\begin{figure}[t!]
    \centering
\[\Qcircuit @C=1.7em @R=1em {    \lstick{} & \push{(x,a)\ \ \ } & \multigate{1}{\text{BS}}     &\gate{P(-2\Delta t)}&\qw&\link{1}{-1}  & \multigate{1}{\text{BS}} &\qw &\link{1}{-1}&\qw\\
\lstick{} & \push{(x+1,a)\ \ \ } &\ghost{\text{BS}}  & \qw&\qw&\link{-1}{-1}& \ghost{\text{BS}} &\qw &\link{-1}{-1} &\qw   }  \]
    \caption{Quantum circuit for a Trotter step of the interaction term (Eq.\ \eqref{eq:427}) in the Hamiltonian \eqref{eq:og_H} for qumodes $\phi_a (x)$ and $\phi_a (x+1)$.}
    \label{fig:4aaa}
\end{figure}
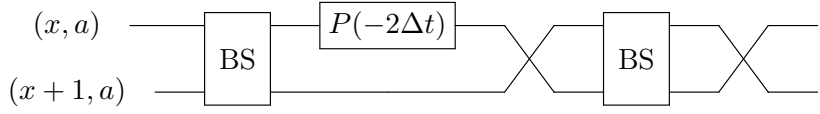

\begin{figure}[ht!]
    \centering
\[ \ \ \ \ \ \ \ \ \  \ \Qcircuit @C=1.7em @R=1em {    \lstick{} & \push{(x,1)\ \ \ } & \multigate{1}{U}     &\qw  & \qw& \link{1}{-1}  & \qw &\qw&\qw &\qw &\link{1}{-1} &\qw \\
\lstick{} & \push{(x,2)\ \ \ } &\ghost{U}  & \multigate{1}{U} & \qw& \link{-1}{-1} &\link{1}{-1} & \multigate{1}{U} &\qw &\link{1}{-1} &\link{-1}{-1} &\qw \\
\lstick{} & \push{(x,3)\ \ \ } &\qw &\ghost{U}  & \qw  &\qw   &\link{-1}{-1} & \ghost{U} &\qw &\link{-1}{-1} &\qw &\qw }  \]
\\
\[\Qcircuit @C=1.7em @R=1em {    \lstick{} & \push{(x,a)\ \ } & \gate{F} & \multigate{1}{\text{BS}}     &\gate{K(-\frac{\Delta t}{2g^2})}  & \multigate{1}{CK(\frac{\Delta t}{g^2})} & \qw & \link{1}{-1} & \multigate{1}{\text{BS}} &\qw&\link{1}{-1}&\gate{F^\dagger}&\qw\\
\lstick{} & \push{(x,b)\ \ } &\qw &\ghost{\text{BS}}  & \gate{K(-\frac{\Delta t}{2g^2})} & \ghost{CK(\frac{\Delta t}{g^2})} &\qw & \link{-1}{-1} & \ghost{\text{BS}}  &\qw&\link{-1}{-1}&\qw &\qw\inputgroup{1}{2}{.75em}{U_{ab} (x):} }  \]
\caption{Quantum circuit for a Trotter step of the kinetic energy term \eqref{eq:428} of the Hamiltonian \eqref{eq:og_H} for the three qumodes $\phi_a (x)$.}
    \label{fig:4aab}
\end{figure}
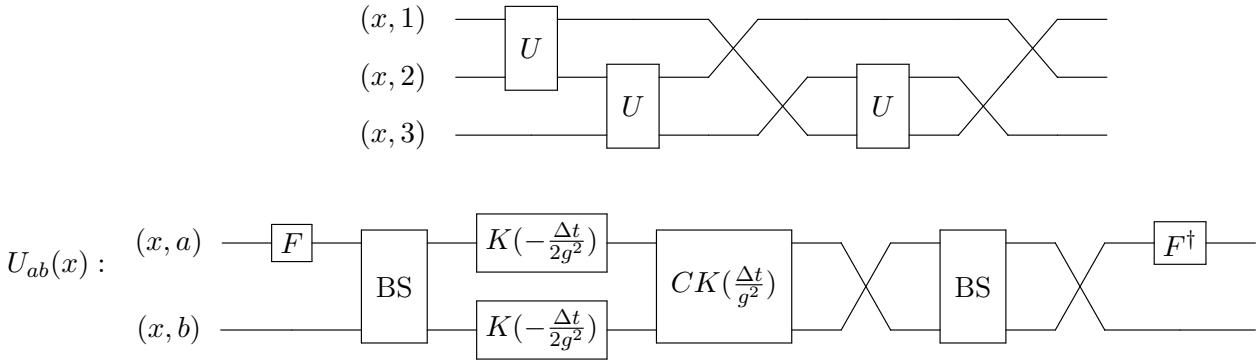

In this section, we perform the time evolution for the Hamiltonian given 
in \eqref{eq:og_H} in terms of small time steps $\Delta t$. In order to promote the interaction term of the Hamiltonian \eqref{eq:og_H} to a time-evolution operator, we apply the string of unitaries \be U_I \equiv \prod_x \prod_a U_a(x) \,, \ee 
where $U_a(x)$ is obtained by applying 50:50 beam splitters and a quadratic phase gate
\be\label{eq:427} U_a (x) \equiv e^{-i\frac{\Delta t}{2}\left( {\phi}_a(x)- {\phi}_a(x+1) \right)^2} = BS_{x+1,x}^a \cdot P_{x,a} \Big( -2\Delta t\Big) \cdot BS_{x,x+1}^a \,.\ee
A quantum circuit implementing this Trotter step is shown in Fig. \ref{fig:4aaa} for qumodes $\phi_a (x)$ and $\phi_a (x+1)$.
The time evolution of the kinetic term can be implemented with the aid of non-Gaussian Kerr gates, $K(s) = e^{is N^{2}}$, and Cross-Kerr gates, 
$CK(s) = e^{is N_{1}N_{2}}$. For a given lattice site $x$, each Trotter step will contain the following string of unitaries
\be\label{eq:428} U_{31} (x) \cdot U_{23} (x) \cdot U_{12} (x)\,. \ee 
Using Eq.\ \eqref{eq:425}, we can write the unitaries as
\be U_{ab} (x) = F_{x,a}^\dagger \cdot BS_x^{ba} \cdot CK_{x}^{ab}\Big(\frac{\Delta t}{g^2}\Big)  \cdot K_{x,a} \Big(-\frac{\Delta t}{2g^2}\Big) \cdot K_{x,b} \Big(-\frac{\Delta t}{2g^2}\Big)  \cdot BS_x^{ab} \cdot F_{x,a} \,. \ee
A quantum circuit implementing the Trotter step for the kinetic energy \eqref{eq:428} is shown in Fig.\ \ref{fig:4aab} for the three qumodes at a given lattice site $x$.

\begin{figure}
    \centering
    \includegraphics[scale=0.55]{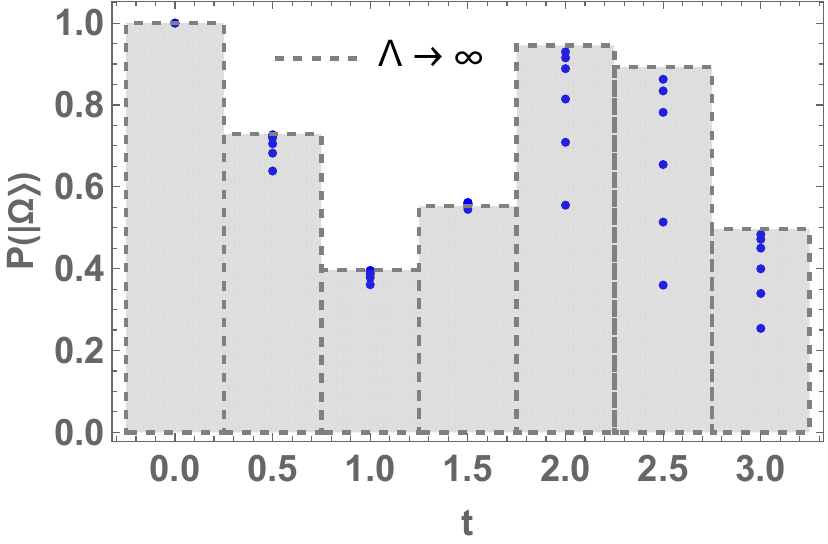}
    \caption{The probability of obtaining the original state $\ket{\Omega(\Lambda)}$ (return probability) as a function of time $t$ for $L=2,l_\text{max.}=3,g^2=1$. The blue points correspond to $\Lambda=3.2,4.5,6.3,10,14,20$ that converge to the $\text{O(3)}$ result shown by the horizontal dashed line in the large $\Lambda$ limit.}
    \label{fig:time_ev_gsq1_lmax3_t2}
\end{figure}

Consider the state $\ket{l,m;\Lambda}$ given by Eq.\ \eqref{eq:basis}. Its wave function factorizes into radial and angular parts
\be \braket{r,\bm{n} \vert l,m;\Lambda} = \psi_{\Lambda} ({r}) Y_{lm} (\bm{n} ) \ , \ \ \psi_{\Lambda} ({r}) = \frac{1}{\sqrt{\mathcal{N}}}   e^{-\Lambda^2\left(r^2-g^2\right)^2/8g^2} \,,\ee 
where the normalization constant is given in \eqref{eq:35} for large values of the cutoff $\Lambda$.
The radial wave function is centered at $r=g$ with a spread $\Delta r \sim \mathcal{O} (1/\Lambda)$. 
The matrix elements of the evolution operator in the basis of \eqref{eq:basis} can be written as
\be \label{eq:534}\bra{\bm{l},\bm{m};\Lambda} e^{-itH} \ket{\bm{l}',\bm{m}';\Lambda} = \int \prod_x dr(x) r^2(x) \, |\psi_\Lambda (r(x))|^2 \bra{\bm{lm}} e^{-itH} \ket{\bm{l}'\bm{m}'}~. \ee 
Here $\ket{\bm{l},\bm{m};\Lambda} = \bigotimes_x \ket{l(x),m(x);\Lambda}$, and similarly for the other states.
The Hamiltonian $H$ is a function of the radial coordinates through its interaction term which  is quadratic in the fields. Therefore, the radial spread in the exponent of the evolution operator is $t\cdot \mathcal{O} (r\Delta r) \sim \mathcal{O} (gt/\Lambda)$, where we used $r\sim g$ and $\Delta r \sim 1/\Lambda$.
It follows that, given $t$, we need to choose $\Lambda$ so that $t\lesssim \Lambda /g$.
To show this numerically, we computed the probability corresponding to the transition amplitudes in \eqref{eq:534} for $g^2 =1$ and 
several values of $\Lambda$. The results are shown in Fig.~\ref{fig:time_ev_gsq1_lmax3_t2}. 

In this Figure, we computed this probability for $l=l'=m=m'=0$ and $L=2$. As in \eqref{eq:534}, this calculation was performed by expressing the time evolution operator as a matrix in the basis of spherical harmonics, and as a function of the continuous parameter $r$. Monte Carlo integration was then performed over the $r$ coordinate to complete the computation of the amplitude. We find that greater values of $\Lambda$ are required for large times $t$ as well as for values of $t$ for which the probability of returning to the original state is large. 

\begin{figure}
    \centering
    \includegraphics[scale=0.5]{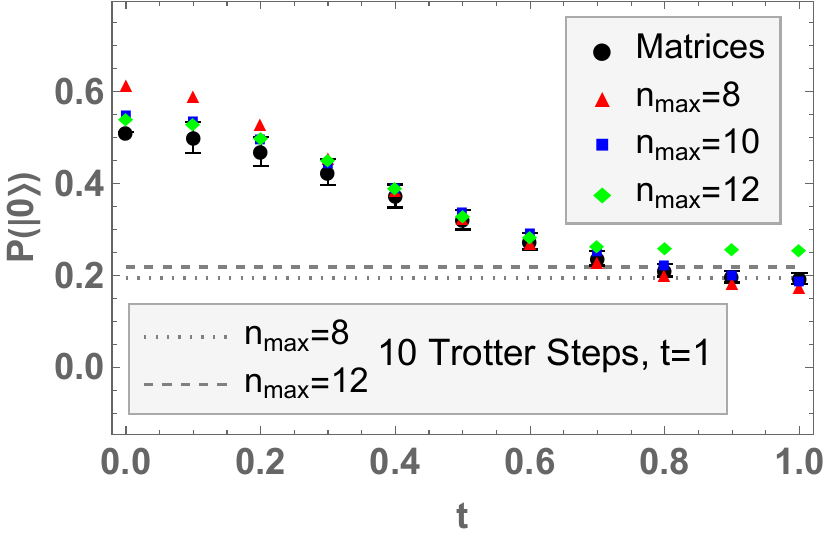}
    \caption{Probability of obtaining the vacuum state $\ket{\textbf{0}}$, after evolving the state $\ket{\Omega(\Lambda)}$ for time $t$ with $L=2,g^2=1,\Lambda= 3.2$. The black points are obtained in the same way as in Fig. \ref{fig:time_ev_gsq1_lmax3_t2} (matrices in spherical harmonic basis with truncation $l_\text{max.}=3$, Monte Carlo integration over the radial direction). The points without error bars indicate results obtained with the quantum simulator \SF~ with two Trotter steps for each time measurement, for different Fock space cutoffs $n_\text{max}$. We also show the result for $n_\text{max}=8$ and $n_\text{max}=12$ with $10$ Trotter steps, for $t=1$.}
    \label{fig:TE_gsq1_R10}
\end{figure}

In Fig.~\ref{fig:TE_gsq1_R10}, we show results for the time evolution using the \SF\ quantum simulator. Instead of computing the amplitude in the basis $\ket{\bm{l},\bm{m};\Lambda}$, we constructed and evolved the state $\ket{\bm{0},\bm{0};\Lambda}$ but computed the overlap with the photon number vacuum state
\be \langle\bm r\vert\bm 0\rangle\propto e^{-\bm r^2/2}\ket{\Omega_0}\ee
where $\ket{\Omega_0}$ is defined in Eq.\ \eqref{eq:CCOmega0} and $\bm{r}$ is the vector of radial coordinates over all sites. Even with just two Trotter steps for each time measurement (colored markers) and a small truncation in the photon number basis, the results in 
Fig.~\ref{fig:TE_gsq1_R10} display good agreement with results obtained using matrices and Monte Carlo integration, the latter of which contains no Trotter error.

\section{Conclusion~\label{sec:conclu}}

We studied the $\text{O(3)}$ nonlinear sigma model in 1+1 dimensions using continuous variable quantum computing. Instead of using discrete variables (qubits), continuous variables (qumodes) allow for an infinite dimensional Hilbert space, which is well suited for models involving bosonic degrees of freedom. To achieve this, instead of following the Schwinger boson approach based on two qumodes per site \cite{Jha:2023ump,Davoudi:2022xmb}, we considered a collection of scalar fields with three physical qumodes for each lattice site. We considered wave functions that are peaked on the 2D sphere with a radial spread defined by a momentum cutoff $\Lambda$. In the limit $\Lambda\to\infty$, we showed that our results agree with the lattice rotor Hamiltonian for the nonlinear $\text{O(3)}$ sigma model. We carried out the quantum simulation by calculating matrix elements of the Hamiltonian in a truncated Hilbert space and then obtained energy levels by diagonalizing the resulting matrix. We compared these results to an approach based on a Coupled-Cluster Ansatz for the ground and excited states and found that the latter accurately yielded the ground and excited state energies. Using this Ansatz, we outlined the quantum algorithm to compute energy levels and perform time evolution of the system using continuous variables. To show that our outlined procedure works in practice, we prepared the ground state Ansatz and performed time evolution calculations using the photonic simulator \SF~with an appropriate truncation of the Fock space. 

Aside from the $\text{O(3)}$ model being interesting in its own right, an improved understanding of the nonlinear sigma model represents a step toward addressing some of the most challenging open questions in fundamental nuclear and particle physics. This includes real-time dynamics relevant to inelastic scattering processes at collider experiments, hadron structure, QCD hadronization, and the dynamics of nuclear matter under extreme conditions. The $\text{O(3)}$ model shares similarities with four-dimensional QCD including the presence of instanton solutions, asymptotic freedom, and the presence of a dynamically generated mass gap. One of the critical next steps will be to promote the global symmetry discussed in this work to local gauge symmetries using the continuous variable approach to quantum computing, which we will address in future work. We expect that our work will facilitate further exploration of models with increasing complexity eventually leading to an improved understanding of QCD. 

\acknowledgments
We would like to thank Jack Araz, Ra\'ul Brice\~no, Zohreh Davoudi, Robert Edwards, and Kostas Orginos for helpful discussions. The research was supported by the U.S. Department of Energy, Office of Science, National Quantum Information Science Research Centers, Co-design Center for Quantum Advantage under contract number DE-SC0012704 and the U.S. Department of Energy, Office of Science, Office of Nuclear Physics under contract DE-AC05-06OR23177. GS and ST acknowledge support by DOE ASCR funding under the Quantum Computing Application Teams Program, the Army Research Office award W911NF-19-1-0397, NSF award DGE-2152168, and DOE, Office of Nuclear Physics, Quantum Horizons Program award DE-SC0023687. RGJ and FR are supported by the U.S. Department of Energy, Office of Science, Contract No.~DE-AC05-06OR23177, under which Jefferson Science Associates, LLC operates Jefferson Lab. FR is supported in part by the DOE, Office of Science, Office of Nuclear Physics, Early Career Program under contract No. DE-SC0024358. This research used resources of the Oak Ridge Leadership Computing Facility, which is a DOE Office of Science User Facility supported under Contract DE-AC05-00OR22725.

\appendix
\section{Continuous-variable gates~\label{sec:CV_gates}}

In this appendix, we review some of the CV gates used in this paper. Additional details and gates can be found in \cite{Killoran_2019}. 
\subsection{Single qumode gates}
To create a squeezed state on the sphere, one must first apply a squeeze operation and then a displacement in position. Squeezing is implemented by 
\be\label{eq:Sdef} S(r)=e^{\frac{r}{2}(a^{\dagger^2}-a^2)},\ee
where $a$ and $a^\dagger$ are bosonic creation and annihilation operators with $[a,a^\dagger]=1$. The position and conjugate momentum are written as
\be \label{eq:qpaadag} q=\frac{1}{\sqrt2}(a^\dagger + a)\ ,\ \ p=\frac{i}{\sqrt2}(a^\dagger - a) \,. \ee
Recall that in this work the components of the field $\phi_i(x)$ at $L$ sites, give $3L$ independent position operators, each of which are accompanied by their conjugate momentum $\pi_i(x)$. It follows from \eqref{eq:Sdef} and \eqref{eq:qpaadag} that
\be S^\dagger(r)qS(r)=e^{-r}q\ ,\ \ S^\dagger(r)pS(r)=e^{r}p  \,.\ee
Note that we have 
\be r=\ln \Lambda,\ee
where $\Lambda$ is the cutoff parameter restricting the wave functions to the sphere. A displacement in position can be achieved using a displacement gate, which depends on a real-valued parameter $x\in \mathbb{R}$ as
\be e^{-ipx}=D(x/\sqrt2)=e^{\frac{x}{\sqrt2} \left(a^\dagger -a\right)}\,.\ee
Its action is given by
\be q\to q+x\ ,\ \ p\to p \,. \ee
This action gives us a way to compute the expectation value of $q$ as
\be  q= \frac{1}{2x}\left(e^{ipx}Ne^{-ipx}-e^{-ipx}Ne^{ipx}\right)\ , \ \ N=\frac{1}{2}\left(q^2+p^2\right) \,. \ee
Thus to compute $\bra{\psi}q\ket{\psi}$, we compute the mean photon number in the two states
\be e^{-ipx}\ket{\psi}\ ,\ \ e^{ipx}\ket{\psi} \ee
and take the difference. 

Next, the rotation gate is defined by
\be\label{eq:R} R(\theta)=e^{iN\theta}\ ,\ \ \theta\in \mathbb{R} \,.\ee
It rotates the position and momentum as
\be \begin{pmatrix} q \\ p \end{pmatrix}\to\begin{pmatrix}
    \cos\theta & -\sin\theta \\ \sin\theta & \cos\theta \end{pmatrix}\begin{pmatrix}q \\ p\end{pmatrix} \,.\ee
Another useful gate we have used in the main text is the Fourier transform $F$. This gate is the continuous variable version of the Hadamard gate. It is a special case of the rotation gate
\begin{equation}\label{eq:aF}
    F = R\left(\frac{\pi}{2}\right) = e^{\frac{i\pi}{4}(p^{2} + q^{2})} \,.
\end{equation}
We make use of it in creating the state $\ket{\Omega(\Lambda)}$, see Eq.~\eqref{eq:Omega_def}, and to help implement the angular momentum operator $\bm{L}$. 

A particularly useful single-qumode gate is the quadratic phase gate,
\be\label{eq:aP} P(s)=e^{isq^2/2}, \ee
which has the transformation properties
\be P^\dagger(s)qP(s)= q\ ,\ \ P^\dagger(s)pP(s)= p+sq\,.\ee
From this, we find
\be P^\dagger (s)NP(s)=P^\dagger (s)\frac{1}{2}\left(q^2+p^2\right)P(s)=\frac{1}{2}\left[q^2+\left(p+sq\right)^2\right]\,,\ee
and thus 
\be q^2=\frac{1}{s^2}\left(P^\dagger(s)NP(s)+P^\dagger(-s)NP(-s)-2P^\dagger(0)NP(0)\right) \,.\ee
This means that we can compute the expectation value $\bra{\psi}q^2\ket{\psi}$ by instead computing
\be \bra{\psi}P^\dagger (s) NP(s)\ket{\psi},\ee
for three values of $s$. These are found simply by measuring the mean photon number in the state $P(s)\ket{\psi}$. Note that it is possible to express the $P$ gate in terms of the more elementary rotation and single-mode squeeze gates \cite{Killoran_2019}.

\subsection{Two qumode gates}

The fundamental two qumode gate that is typically considered is the beamsplitter gate given by
\be \text{BS}_{ab}(\theta) = e^{\theta(a b^\dagger - a^\dagger b)}=\text{BS}_{ba}^\dagger(\theta),\ \theta\in \mathbb{R} \,.
\label{eq:BS_def}
\ee
We use it in conjunction with $F$ to implement the angular momentum operator $\bm{L}$. Its action is given by
\be \begin{pmatrix} a \\ b \end{pmatrix}\to\begin{pmatrix}
    \cos\theta & -\sin\theta \\ \sin\theta & \cos\theta \end{pmatrix}\begin{pmatrix} a \\ b\end{pmatrix}\,.\ee
We note that in our qumode formulation, if $a$ and $b$ belong to two of the components of a triplet $\bm{q}$, then the beam splitter corresponds to an $\text{O(3)}$ rotation in the $ab$-plane
\be \text{BS}_{ab} (\theta)=e^{-i\theta J_{ab}}\,. \ee
Another important gate is the CX gate, which is given by
\be\label{eq:aCX} \text{CX}_{ab}(s)=e^{-is q_a p_b}\,. \ee
It can be decomposed in terms of beam splitters and single-mode squeezers. Its action is
\bea
&& q_a\to q_a\ ,\ \quad p_a\to p_a-sp_b\,,\nonumber\\
&&  p_b\to p_b\ ,\ \quad q_b\to q_b+sq_a\,.
\eea
Among other use cases, it enables us to engineer certain exponential wave functions with the aid of an ancilla. Let the quadratures $(q_b,p_b)$ represent a physical qumode and $(q_a,p_a)$ an ancilla. Then 
\be \text{CX}_{ba}(s)\ket{\psi}_b\otimes\ket{0}_a=\int dq_b \langle q_b\vert\psi\rangle\ket{q_b}_b\otimes e^{-iap_aq_b}\ket{0}_a=\int dq_b \langle q_b\vert\psi\rangle\ket{q}_b\otimes \ket{sq_b}_a \,. \ee
The state $\ket{sq_b}_a$ is a coherent state, and so 
\be \langle N=0\vert sq_b\rangle_a = e^{-s^2q_b^2/2}\,. \ee
Thus, measuring $q_a$ to be in the $N=0$ state gives the (unnormalized) state
\be {}_a\bra{N=0} \text{CX}_{ba}(s)\ket{\psi}_b\otimes\ket{0}_a = e^{-s^2q_b^2/2}\ket{\psi}_b \,. \ee
This recipe allows us to apply the coupled-cluster operator to the state $\ket{\Omega(\Lambda)}$. In this case, we couple two physical modes to an ancilla at a time:
\be q_bp_a \to \left( \phi_b (x)- \phi_b (x+1)\right)p_a \ , \ \ b=1,2,3\ . \ee
A similar procedure is used to construct $\ket{\Omega(\Lambda)}$ itself. In this case, we make use of a similar operation called the CZ gate, defined by
\be \text{CZ}_{ab}(s)=e^{is q_a q_b} \ . \ee
The CZ gate is related to the CX gate via the Fourier transform operator \eqref{eq:aF}
as $\text{CZ}_{ab} = F_b^\dagger \cdot \text{CX}_{ab} \cdot F_b$.
However, additional elements are required since the operation is non-Gaussian. We refer the reader to Ref.~\cite{Killoran_2019} for information on additional gates, such as the cubic phase and Kerr gates, both of which implement non-Gaussian operations.

\bibliography{main.bib}

\providecommand{\href}[2]{#2}\begingroup\raggedright\begin{thebibliography}{10}

\bibitem{RevModPhys.93.045003}
J.~I. Cirac, D.~P\'erez-Garc\'{\i}a, N.~Schuch, and F.~Verstraete, ``Matrix product states and projected entangled pair states: Concepts, symmetries, theorems,'' \href{http://dx.doi.org/10.1103/RevModPhys.93.045003}{{\em Rev. Mod. Phys.} {\bfseries 93} (Dec, 2021) 045003}. \url{https://link.aps.org/doi/10.1103/RevModPhys.93.045003}.

\bibitem{Bauer:2022hpo}
C.~W. Bauer {\em et~al.}, ``{Quantum Simulation for High-Energy Physics},'' \href{http://dx.doi.org/10.1103/PRXQuantum.4.027001}{{\em PRX Quantum} {\bfseries 4} no.~2, (2023) 027001}, \href{http://arxiv.org/abs/2204.03381}{{\ttfamily arXiv:2204.03381 [quant-ph]}}.

\bibitem{Briceno:2020rar}
R.~A. Brice\~no, J.~V. Guerrero, M.~T. Hansen, and A.~M. Sturzu, ``{Role of boundary conditions in quantum computations of scattering observables},'' \href{http://dx.doi.org/10.1103/PhysRevD.103.014506}{{\em Phys. Rev. D} {\bfseries 103} no.~1, (2021) 014506}, \href{http://arxiv.org/abs/2007.01155}{{\ttfamily arXiv:2007.01155 [hep-lat]}}.

\bibitem{Polyakov:1975rr}
A.~M. Polyakov, ``{Interaction of Goldstone Particles in Two-Dimensions. Applications to Ferromagnets and Massive Yang-Mills Fields},'' \href{http://dx.doi.org/10.1016/0370-2693(75)90161-6}{{\em Phys. Lett. B} {\bfseries 59} (1975) 79--81}.

\bibitem{Polyakov:1975yp}
A.~M. Polyakov and A.~A. Belavin, ``{Metastable States of Two-Dimensional Isotropic Ferromagnets},'' {\em JETP Lett.} {\bfseries 22} (1975) 245--248. \url{http://jetpletters.ru/ps/1529/article_23383.shtml}.

\bibitem{Berg:1981er}
B.~Berg and M.~Luscher, ``{Definition and Statistical Distributions of a Topological Number in the Lattice O(3) Sigma Model},'' \href{http://dx.doi.org/10.1016/0550-3213(81)90568-X}{{\em Nucl. Phys. B} {\bfseries 190} (1981) 412--424}.

\bibitem{Bruckmann:2018usp}
F.~Bruckmann, K.~Jansen, and S.~K\"uhn, ``{O(3) nonlinear sigma model in 1+1 dimensions with matrix product states},'' \href{http://dx.doi.org/10.1103/PhysRevD.99.074501}{{\em Phys. Rev. D} {\bfseries 99} no.~7, (2019) 074501}, \href{http://arxiv.org/abs/1812.00944}{{\ttfamily arXiv:1812.00944 [hep-lat]}}.

\bibitem{Tang:2021uge}
W.~Tang, X.~C. Xie, L.~Wang, and H.-H. Tu, ``{Tensor network simulation of the (1+1)-dimensional O(3) nonlinear \ensuremath{\sigma}-model with \ensuremath{\theta}=\ensuremath{\pi} term},'' \href{http://dx.doi.org/10.1103/PhysRevD.104.114513}{{\em Phys. Rev. D} {\bfseries 104} no.~11, (2021) 114513}, \href{http://arxiv.org/abs/2109.11324}{{\ttfamily arXiv:2109.11324 [hep-lat]}}.

\bibitem{Unmuth-Yockey:2014afa}
J.~F. Unmuth-Yockey, Y.~Meurice, J.~Osborn, and H.~Zou, ``{Tensor renormalization group study of the 2d O(3) model},'' \href{http://dx.doi.org/10.22323/1.214.0325}{{\em PoS} {\bfseries LATTICE2014} (2014) 325}, \href{http://arxiv.org/abs/1411.4213}{{\ttfamily arXiv:1411.4213 [hep-lat]}}.

\bibitem{Brower:2003vy}
R.~Brower, S.~Chandrasekharan, S.~Riederer, and U.~J. Wiese, ``{D theory: Field quantization by dimensional reduction of discrete variables},'' \href{http://dx.doi.org/10.1016/j.nuclphysb.2004.06.007}{{\em Nucl. Phys. B} {\bfseries 693} (2004) 149--175}, \href{http://arxiv.org/abs/hep-lat/0309182}{{\ttfamily arXiv:hep-lat/0309182}}.

\bibitem{Chandrasekharan:2001ya}
S.~Chandrasekharan, B.~Scarlet, and U.~J. Wiese, ``{From spin ladders to the 2-d O(3) model at nonzero density},'' \href{http://dx.doi.org/10.1016/S0010-4655(02)00311-9}{{\em Comput. Phys. Commun.} {\bfseries 147} (2002) 388--393}, \href{http://arxiv.org/abs/hep-lat/0110215}{{\ttfamily arXiv:hep-lat/0110215}}.

\bibitem{Singh:2019jog}
H.~Singh, ``{Qubit regularized O(N) nonlinear sigma models},'' \href{http://dx.doi.org/10.1103/PhysRevD.105.114509}{{\em Phys. Rev. D} {\bfseries 105} no.~11, (2022) 114509}, \href{http://arxiv.org/abs/1911.12353}{{\ttfamily arXiv:1911.12353 [hep-lat]}}.

\bibitem{Alexandru:2019ozf}
{\bfseries NuQS} Collaboration, A.~Alexandru, P.~F. Bedaque, H.~Lamm, and S.~Lawrence, ``{\ensuremath{\sigma} Models on Quantum Computers},'' \href{http://dx.doi.org/10.1103/PhysRevLett.123.090501}{{\em Phys. Rev. Lett.} {\bfseries 123} no.~9, (2019) 090501}, \href{http://arxiv.org/abs/1903.06577}{{\ttfamily arXiv:1903.06577 [hep-lat]}}.

\bibitem{Buser:2020uzs}
A.~J. Buser, T.~Bhattacharya, L.~Cincio, and R.~Gupta, ``{State preparation and measurement in a quantum simulation of the $O$(3) sigma model},'' \href{http://dx.doi.org/10.1103/PhysRevD.102.114514}{{\em Phys. Rev. D} {\bfseries 102} no.~11, (2020) 114514}, \href{http://arxiv.org/abs/2006.15746}{{\ttfamily arXiv:2006.15746 [quant-ph]}}.

\bibitem{Caspar:2022llo}
S.~Caspar and H.~Singh, ``{From Asymptotic Freedom to \ensuremath{\theta} Vacua: Qubit Embeddings of the O(3) Nonlinear \ensuremath{\sigma} Model},'' \href{http://dx.doi.org/10.1103/PhysRevLett.129.022003}{{\em Phys. Rev. Lett.} {\bfseries 129} no.~2, (2022) 022003}, \href{http://arxiv.org/abs/2203.15766}{{\ttfamily arXiv:2203.15766 [hep-lat]}}.

\bibitem{Araz:2022tbd}
J.~Y. Araz, S.~Schenk, and M.~Spannowsky, ``Toward a quantum simulation of nonlinear sigma models with a topological term,'' \href{http://dx.doi.org/10.1103/PhysRevA.107.032619}{{\em Phys. Rev. A} {\bfseries 107} (Mar, 2023) 032619}. \url{https://link.aps.org/doi/10.1103/PhysRevA.107.032619}.

\bibitem{Alexandru:2022son}
A.~Alexandru, P.~F. Bedaque, A.~Carosso, M.~J. Cervia, and A.~Sheng, ``{Qubitization strategies for bosonic field theories},'' \href{http://dx.doi.org/10.1103/PhysRevD.107.034503}{{\em Phys. Rev. D} {\bfseries 107} no.~3, (2023) 034503}, \href{http://arxiv.org/abs/2209.00098}{{\ttfamily arXiv:2209.00098 [hep-lat]}}.

\bibitem{Jha:2023ump}
R.~Jha, F.~Ringer, G.~Siopsis, and S.~Thompson, \href{http://dx.doi.org/10.22323/1.453.0230}{``Quantum computations of the o(3) model using qumodes,''} in {\em Proceedings of The 40th International Symposium on Lattice Field Theory — PoS(LATTICE2023)}, LATTICE2023.
\newblock Sissa Medialab, Jan., 2024.
\newblock \url{http://dx.doi.org/10.22323/1.453.0230}.

\bibitem{Ciavarella:2022qdx}
A.~N. Ciavarella, S.~Caspar, H.~Singh, and M.~J. Savage, ``{Preparation for quantum simulation of the (1+1)-dimensional O(3) nonlinear \ensuremath{\sigma} model using cold atoms},'' \href{http://dx.doi.org/10.1103/PhysRevA.107.042404}{{\em Phys. Rev. A} {\bfseries 107} no.~4, (2023) 042404}, \href{http://arxiv.org/abs/2211.07684}{{\ttfamily arXiv:2211.07684 [quant-ph]}}.

\bibitem{Singh:2019uwd}
H.~Singh and S.~Chandrasekharan, ``{Qubit regularization of the $O(3)$ sigma model},'' \href{http://dx.doi.org/10.1103/PhysRevD.100.054505}{{\em Phys. Rev. D} {\bfseries 100} no.~5, (2019) 054505}, \href{http://arxiv.org/abs/1905.13204}{{\ttfamily arXiv:1905.13204 [hep-lat]}}.

\bibitem{Alexandru:2021xkf}
A.~Alexandru, P.~F. Bedaque, A.~Carosso, and A.~Sheng, ``{Universality of a truncated sigma-model},'' \href{http://dx.doi.org/10.1016/j.physletb.2022.137230}{{\em Phys. Lett. B} {\bfseries 832} (2022) 137230}, \href{http://arxiv.org/abs/2109.07500}{{\ttfamily arXiv:2109.07500 [hep-lat]}}.

\bibitem{Jordan:2011ci}
S.~P. Jordan, K.~S.~M. Lee, and J.~Preskill, ``{Quantum Computation of Scattering in Scalar Quantum Field Theories},'' {\em Quant. Inf. Comput.} {\bfseries 14} (2014) 1014--1080, \href{http://arxiv.org/abs/1112.4833}{{\ttfamily arXiv:1112.4833 [hep-th]}}.

\bibitem{Marshall:2015mna}
K.~Marshall, R.~Pooser, G.~Siopsis, and C.~Weedbrook, ``{Quantum simulation of quantum field theory using continuous variables},'' \href{http://dx.doi.org/10.1103/PhysRevA.92.063825}{{\em Phys. Rev. A} {\bfseries 92} no.~6, (2015) 063825}, \href{http://arxiv.org/abs/1503.08121}{{\ttfamily arXiv:1503.08121 [quant-ph]}}.

\bibitem{Yeter2022}
K.~Yeter-Aydeniz, E.~Moschandreou, and G.~Siopsis, ``Quantum imaginary-time evolution algorithm for quantum field theories with continuous variables,'' \href{http://dx.doi.org/10.1103/PhysRevA.105.012412}{{\em Phys. Rev. A} {\bfseries 105} (Jan, 2022) 012412}. \url{https://link.aps.org/doi/10.1103/PhysRevA.105.012412}.

\bibitem{PhysRevLett.82.1784}
S.~Lloyd and S.~L. Braunstein, ``Quantum computation over continuous variables,'' \href{http://dx.doi.org/10.1103/PhysRevLett.82.1784}{{\em Phys. Rev. Lett.} {\bfseries 82} (Feb, 1999) 1784--1787}. \url{https://link.aps.org/doi/10.1103/PhysRevLett.82.1784}.

\bibitem{PhysRevLett.127.020502}
D.-B. Zhang, G.-Q. Zhang, Z.-Y. Xue, S.-L. Zhu, and Z.~D. Wang, ``Continuous-variable assisted thermal quantum simulation,'' \href{http://dx.doi.org/10.1103/PhysRevLett.127.020502}{{\em Phys. Rev. Lett.} {\bfseries 127} (Jul, 2021) 020502}. \url{https://link.aps.org/doi/10.1103/PhysRevLett.127.020502}.

\bibitem{PhysRevLett.97.110501}
N.~C. Menicucci, P.~van Loock, M.~Gu, C.~Weedbrook, T.~C. Ralph, and M.~A. Nielsen, ``Universal quantum computation with continuous-variable cluster states,'' \href{http://dx.doi.org/10.1103/PhysRevLett.97.110501}{{\em Phys. Rev. Lett.} {\bfseries 97} (Sep, 2006) 110501}. \url{https://link.aps.org/doi/10.1103/PhysRevLett.97.110501}.

\bibitem{Xanadu22}
L.~S. Madsen, F.~Laudenbach, M.~F. Askarani, F.~Rortais, T.~Vincent, J.~F.~F. Bulmer, F.~M. Miatto, L.~Neuhaus, L.~G. Helt, M.~J. Collins, A.~E. Lita, T.~Gerrits, S.~W. Nam, V.~D. Vaidya, M.~Menotti, I.~Dhand, Z.~Vernon, N.~Quesada, and J.~Lavoie, ``Quantum computational advantage with a programmable photonic processor,'' \href{http://dx.doi.org/10.1038/s41586-022-04725-x}{{\em Nature} {\bfseries 606} no.~7912, (2022) 75--81}. \url{https://doi.org/10.1038/s41586-022-04725-x}.

\bibitem{Taballione:2022xjq}
C.~Taballione {\em et~al.}, ``{20-Mode Universal Quantum Photonic Processor},'' \href{http://dx.doi.org/10.22331/q-2023-08-01-1071}{{\em Quantum} {\bfseries 7} (2023) 1071}, \href{http://arxiv.org/abs/2203.01801}{{\ttfamily arXiv:2203.01801 [quant-ph]}}.

\bibitem{Eaton:2022vjq}
M.~Eaton, A.~Hossameldin, R.~J. Birrittella, P.~M. Alsing, C.~C. Gerry, H.~Dong, C.~Cuevas, and O.~Pfister, ``{Resolution of 100 photons and quantum generation of unbiased random numbers},'' \href{http://dx.doi.org/10.1038/s41566-022-01105-9}{{\em Nature Photon.} {\bfseries 17} no.~1, (2023) 106--111}, \href{http://arxiv.org/abs/2205.01221}{{\ttfamily arXiv:2205.01221 [quant-ph]}}.

\bibitem{Polyakov:1987ez}
A.~M. Polyakov, ``{Gauge Fields and Strings Vol.~3 (1987)},''.

\bibitem{PhysRevD.11.395}
J.~Kogut and L.~Susskind, ``Hamiltonian formulation of wilson's lattice gauge theories,'' \href{http://dx.doi.org/10.1103/PhysRevD.11.395}{{\em Phys. Rev. D} {\bfseries 11} (Jan, 1975) 395--408}. \url{https://link.aps.org/doi/10.1103/PhysRevD.11.395}.

\bibitem{Hamer:1978ew}
C.~J. Hamer, J.~B. Kogut, and L.~Susskind, ``{Strong Coupling Expansions and Phase Diagrams for the O(2), O(3) and O(4) Heisenberg Spin Systems in Two-dimensions},'' \href{http://dx.doi.org/10.1103/PhysRevD.19.3091}{{\em Phys. Rev. D} {\bfseries 19} (1979) 3091}.

\bibitem{Wu1977}
T.~T. Wu and C.~N. Yang, ``Some properties of monopole harmonics,'' \href{http://dx.doi.org/10.1103/physrevd.16.1018}{{\em Physical Review D} {\bfseries 16} no.~4, (Aug., 1977) 1018--1021}. \url{https://doi.org/10.1103/physrevd.16.1018}.

\bibitem{Crawford2000}
T.~D. Crawford and H.~F. Schaefer~III, {\em An Introduction to Coupled Cluster Theory for Computational Chemists}, \href{http://dx.doi.org/https://doi.org/10.1002/9780470125915.ch2}{pp.~33--136}.
\newblock John Wiley \& Sons, Ltd, 2000.
\newblock \url{https://onlinelibrary.wiley.com/doi/abs/10.1002/9780470125915.ch2}.

\bibitem{Kutzelnigg1991}
W.~Kutzelnigg, ``Error analysis and improvements of coupled-cluster theory,'' \href{http://dx.doi.org/10.1007/bf01117418}{{\em Theoretica Chimica Acta} {\bfseries 80} no.~4-5, (1991) 349--386}. \url{https://doi.org/10.1007/bf01117418}.

\bibitem{Ligterink:1997he}
N.~E. Ligterink, N.~R. Walet, and R.~F. Bishop, ``{A Coupled cluster formulation of Hamiltonian lattice field theory: The nonlinear sigma model},'' \href{http://dx.doi.org/10.1006/aphy.1998.5812}{{\em Annals Phys.} {\bfseries 267} (1998) 97--133}, \href{http://arxiv.org/abs/hep-lat/9712021}{{\ttfamily arXiv:hep-lat/9712021}}.

\bibitem{Ligterink1998}
N.~Ligterink, N.~Walet, and R.~Bishop, ``The ground state of the nonlinear sigma model o(4) in 3+1 dimensions,'' \href{http://dx.doi.org/10.1016/s0920-5632(97)00866-9}{{\em Nuclear Physics B - Proceedings Supplements} {\bfseries 63} no.~1-3, (Apr., 1998) 667--669}. \url{https://doi.org/10.1016/s0920-5632(97)00866-9}.

\bibitem{Ryabinkin2018}
I.~G. Ryabinkin, T.-C. Yen, S.~N. Genin, and A.~F. Izmaylov, ``Qubit coupled cluster method: A systematic approach to quantum chemistry on a quantum computer,'' \href{http://dx.doi.org/10.1021/acs.jctc.8b00932}{{\em Journal of Chemical Theory and Computation} {\bfseries 14} no.~12, (Nov., 2018) 6317--6326}. \url{https://doi.org/10.1021/acs.jctc.8b00932}.

\bibitem{Dumitrescu:2018njn}
E.~F. Dumitrescu, A.~J. McCaskey, G.~Hagen, G.~R. Jansen, T.~D. Morris, T.~Papenbrock, R.~C. Pooser, D.~J. Dean, and P.~Lougovski, ``{Cloud Quantum Computing of an Atomic Nucleus},'' \href{http://dx.doi.org/10.1103/PhysRevLett.120.210501}{{\em Phys. Rev. Lett.} {\bfseries 120} (2018) 210501}, \href{http://arxiv.org/abs/1801.03897}{{\ttfamily arXiv:1801.03897 [quant-ph]}}.

\bibitem{Duncan1985}
A.~Duncan and R.~Roskies, ``Variational estimates for spectra in lattice hamiltonian theories,'' \href{http://dx.doi.org/10.1103/PhysRevD.31.364}{{\em Phys. Rev. D} {\bfseries 31} (Jan, 1985) 364--376}. \url{https://link.aps.org/doi/10.1103/PhysRevD.31.364}.

\bibitem{PhysRevD.32.3277}
A.~Duncan and R.~Roskies, ``Asymptotic scaling in hamiltonian calculations of the o(3) \ensuremath{\sigma} model,'' \href{http://dx.doi.org/10.1103/PhysRevD.32.3277}{{\em Phys. Rev. D} {\bfseries 32} (Dec, 1985) 3277--3281}. \url{https://link.aps.org/doi/10.1103/PhysRevD.32.3277}.

\bibitem{Niederreiter1978}
H.~Niederreiter, ``Quasi-monte carlo methods and pseudo-random numbers,'' {\em \href{https://projecteuclid.org/journals/bulletin-of-the-american-mathematical-society-new-series/volume-84/issue-6/Quasi-Monte-Carlo-methods-and-pseudo-random-numbers/bams/1183541461.full}{Bulletin of the American Mathematical Society}} {\bfseries 84} no.~6, (1978) 957 -- 1041.

\bibitem{Thompson:2023kxz}
S.~Thompson and G.~Siopsis, ``\href{https://link.springer.com/article/10.1007/s11128-023-04149-0}{Quantum computation of phase transition in interacting scalar quantum field theory},'' \href{http://dx.doi.org/10.1007/s11128-023-04149-0}{{\em Quantum Information Processing} {\bfseries 22} (10, 2023) }.

\bibitem{Peruzzo2014}
A.~Peruzzo, J.~McClean, P.~Shadbolt, M.-H. Yung, X.-Q. Zhou, P.~J. Love, A.~Aspuru-Guzik, and J.~L. O’Brien, ``A variational eigenvalue solver on a photonic quantum processor,'' \href{http://dx.doi.org/10.1038/ncomms5213}{{\em Nature Communications} {\bfseries 5} no.~1, (July, 2014) }. \url{http://dx.doi.org/10.1038/ncomms5213}.

\bibitem{Yanagimoto2020}
R.~Yanagimoto, T.~Onodera, E.~Ng, L.~G. Wright, P.~L. McMahon, and H.~Mabuchi, ``Engineering a kerr-based deterministic cubic phase gate via gaussian operations,'' \href{http://dx.doi.org/10.1103/PhysRevLett.124.240503}{{\em Phys. Rev. Lett.} {\bfseries 124} (Jun, 2020) 240503}. \url{https://link.aps.org/doi/10.1103/PhysRevLett.124.240503}.

\bibitem{Bharti2022}
K.~Bharti, A.~Cervera-Lierta, T.~H. Kyaw, T.~Haug, S.~Alperin-Lea, A.~Anand, M.~Degroote, H.~Heimonen, J.~S. Kottmann, T.~Menke, W.-K. Mok, S.~Sim, L.-C. Kwek, and A.~Aspuru-Guzik, ``Noisy intermediate-scale quantum algorithms,'' \href{http://dx.doi.org/10.1103/RevModPhys.94.015004}{{\em Rev. Mod. Phys.} {\bfseries 94} (Feb, 2022) 015004}. \url{https://link.aps.org/doi/10.1103/RevModPhys.94.015004}.

\bibitem{Hempel2018}
C.~Hempel, C.~Maier, J.~Romero, J.~McClean, T.~Monz, H.~Shen, P.~Jurcevic, B.~P. Lanyon, P.~Love, R.~Babbush, A.~Aspuru-Guzik, R.~Blatt, and C.~F. Roos, ``Quantum chemistry calculations on a trapped-ion quantum simulator,'' \href{http://dx.doi.org/10.1103/PhysRevX.8.031022}{{\em Phys. Rev. X} {\bfseries 8} (Jul, 2018) 031022}. \url{https://link.aps.org/doi/10.1103/PhysRevX.8.031022}.

\bibitem{Xu2020}
L.~Xu, J.~T. Lee, and J.~K. Freericks, ``Test of the unitary coupled-cluster variational quantum eigensolver for a simple strongly correlated condensed-matter system,'' \href{http://dx.doi.org/10.1142/s0217984920400497}{{\em Modern Physics Letters B} {\bfseries 34} no.~19n20, (Jul, 2020) 2040049}. \url{https://doi.org/10.1142%2Fs0217984920400497}.

\bibitem{Yeter2019}
K.~Yeter-Aydeniz, E.~F. Dumitrescu, A.~J. McCaskey, R.~S. Bennink, R.~C. Pooser, and G.~Siopsis, ``Scalar quantum field theories as a benchmark for near-term quantum computers,'' \href{http://dx.doi.org/10.1103/PhysRevA.99.032306}{{\em Phys. Rev. A} {\bfseries 99} (Mar, 2019) 032306}. \url{https://link.aps.org/doi/10.1103/PhysRevA.99.032306}.

\bibitem{Nelder1965}
J.~A. Nelder and R.~Mead, ``{A Simplex Method for Function Minimization},'' {\em The Computer Journal} {\bfseries 7} no.~4, (01, 1965) 308--313.

\bibitem{Davoudi:2022xmb}
Z.~Davoudi, A.~F. Shaw, and J.~R. Stryker, ``General quantum algorithms for {H}amiltonian simulation with applications to a non-{A}belian lattice gauge theory,'' \href{http://dx.doi.org/10.22331/q-2023-12-20-1213}{{\em {Quantum}} {\bfseries 7} (Dec., 2023) 1213}. \url{https://doi.org/10.22331/q-2023-12-20-1213}.

\bibitem{Killoran_2019}
N.~Killoran, J.~Izaac, N.~Quesada, V.~Bergholm, M.~Amy, and C.~Weedbrook, ``{S}trawberry {F}ields: A software platform for photonic quantum computing,'' \href{http://dx.doi.org/10.22331/q-2019-03-11-129}{{\em Quantum} {\bfseries 3} (2019) 129}, \href{http://arxiv.org/abs/1804.03159}{{\ttfamily arXiv:1804.03159}}.

\end{thebibliography}\endgroup
\bibliographystyle{utphys.bst}

\end{document}